\documentclass[sigconf,authorversion]{acmart}

\AtBeginDocument{%
  }

\copyrightyear{2023}
\acmYear{2023}
\setcopyright{acmlicensed}\acmConference[CHI '23]{Proceedings of the 2023 CHI Conference on Human Factors in Computing Systems}{April 23--28, 2023}{Hamburg, Germany}
\acmBooktitle{Proceedings of the 2023 CHI Conference on Human Factors in Computing Systems (CHI '23), April 23--28, 2023, Hamburg, Germany}
\acmPrice{15.00}
\acmDOI{10.1145/3544548.3580993}
\acmISBN{978-1-4503-9421-5/23/04}

\acmSubmissionID{9438}

\usepackage{acmart-taps}
\usepackage{amsmath}
\usepackage[english]{babel}
\usepackage{bm}
\usepackage{colortbl}
\usepackage{enumitem}
\usepackage{float}
\usepackage[acronym,nohypertypes={acronym},shortcuts,nonumberlist,nolist]{glossaries}

\usepackage{graphicx}
\usepackage{multirow}
\usepackage{subcaption}
\usepackage{tabularx}
\usepackage{xspace}

\addto\extrasenglish{%

}
\microtypecontext{spacing=nonfrench}
\newacronym{AAS}{AAS}{Account Adoption Scale}
\newacronym{ATI}{ATI}{Affinity for Technology Interaction Scale}
\newacronym{BLE}{BLE}{Bluetooth Low Energy}
\newacronym{CTAP}{CTAP}{Client to Authenticator Protocol}
\newacronym[description={Deutsches Forschungsnetz ("German research network")}]{DFN}{DFN}{Deutsches Forschungsnetz}
\newacronym{DNS}{DNS}{Domain Name System}
\newacronym{FIDO}{FIDO}{Fast IDentity Online Alliance}
\newacronym{GDPR}{GDPR}{General Data Protection Regulation}
\newacronym{HTTPS}{HTTPS}{Hypertext Transfer Protocol Secure}
\newacronym{IP}{IP}{Internet Protocol}
\newacronym{ISO}{ISO}{International Organization for Standardization}
\newacronym{NIST}{NIST}{National Institute of Standards and Technology}
\newacronym{OTSP}{OTSP}{Online Token Status Protocol}
\newacronym{PAS}{PAS}{Participant Adoption Scale}
\newacronym{SMS}{SMS}{Short Message Service}
\newacronym{SUS}{SUS}{System Usability Scale}
\newacronym{TLS}{TLS}{Transport Layer Security}
\newacronym{TUDA}{TUDA}{Technische Universität Darmstadt}
\newacronym{U2F}{U2F}{Universal 2nd Factor}
\newacronym{UAF}{UAF}{Universal Authentication Framework}
\newacronym{USB}{USB}{Universal Serial Bus}
\newacronym{W3C}{W3C}{World Wide Web Consortium}
\newacronym{WebAuthn}{WebAuthn}{Web Authentication}

\newacronym[description={One-Factor Authentication}]{1FA}{1FA}{one-factor authentication}
\newacronym[description={Two-Factor Authentication}]{2FA}{2FA}{two-factor authentication}
\newacronym[description={Application Programming Interface}]{API}{API}{application programming interface}
\newacronym[description={Certification Authority}]{CA}{CA}{certification authority}
\newacronym[description={Effect Size}]{ES}{ES}{effect size}
\newacronym[description={False Discovery Rate}]{FDR}{FDR}{false discovery rate}
\newacronym[description={False Rejection Rate}]{FRR}{FRR}{false rejection rate}
\newacronym[description={Graphical User Interface}]{GUI}{GUI}{graphical user interface}
\newacronym[description={Institutional Review Board}]{IRB}{IRB}{institutional review board}
\newacronym[description={Internet Service Provider}]{ISP}{ISP}{internet service provider}
\newacronym[description={Multi-Factor Authentication}]{MFA}{MFA}{multi-factor authentication}
\newacronym[description={Man-in-the-Middle}]{MitM}{MitM}{man-in-the-middle}
\newacronym[description={Near-Field Communication}]{NFC}{NFC}{near-field communication}
\newacronym[description={Operation System}]{OS}{OS}{operating system}
\newacronym[description={One-Time Password}]{OTP}{OTP}{one-time password}
\newacronym[description={Personal Computer}]{PC}{PC}{personal computer}
\newacronym[description={Personal Identification Number}]{PIN}{PIN}{personal identification number}
\newacronym[description={Service Set Identifier}]{SSID}{SSID}{service set identifier}
\newacronym[description={Single Sign-On}]{SSO}{SSO}{single sign-on}
\newacronym[description={Transaction Authentication Number}]{TAN}{TAN}{transaction authentication number}
\newacronym[description={Trusted Platform Module}]{TPM}{TPM}{trusted platform module}
\newacronym[description={Uniform Resource Locator}]{URL}{URL}{uniform resource locator}
\newacronym[description={Wireless Local Area Network}]{WLAN}{WLAN}{wireless local area network}

\newcommand{\altrowcolor}{gray!10}
\newcommand{\briefSection}[1]{\medskip\noindent\textsc{{#1}.}}

\newcommand{\descitem}[1]{\item \textsc{#1}:}
\newcommand{\etal}[1]{{#1} et al.}

\newcommand{\groupP}{Group P\xspace}
\newcommand{\groupR}{Group R\xspace}

\newcommand{\heatcolor}[1]{orange!#1}
\newcommand{\heatmapcell}[1]{\cellcolor{\heatcolor{#1}}{#1\%}}

\newcommand{\kuugel}{\textsc{kuugel}\xspace}

\newcommand{\pnote}{*}
\newcommand{\ppnote}{**}
\newcommand{\pppnote}{***}

\newcommand{\qitem}[1]{({#1})}
\newcommand{\rating}[1]{\noindent$\langle$\textit{{#1}}$\rangle$}

\newcommand{\rotmultirow}[2]{\multirow{#1}*{\rotatebox{90}{#2}}}

\newcommand\VRule[1][\arrayrulewidth]{\vrule width #1}
\newcommand{\studyquote}[3]{{\def\arraystretch{2}\setlength\tabcolsep{7pt}\vspace{1ex} \noindent\begin{tabular}{!{\color{gray!75}\VRule[2pt]}p{\dimexpr\linewidth-2\tabcolsep-0.3pt}}\cellcolor{gray!10}\textit{``#1''} \mbox{{(P#3, #2)}}\tabularnewline\end{tabular}\vspace{1ex}}}

\newcommand{\textquote}[1]{\textit{``#1''}}

\newcommand{\tablefontsize}{\small}
\newcommand{\tablenotesfontsize}{\small}
\newcommand{\tableheadline}{\textbf}

\renewcommand{\emph}{\textit}

\newcommand{\typofix}[1]{\textcolor{\typofixcolor}{#1}}

\newcommand{\typofixcolor}{black}

\newcolumntype{C}[1]{>{\centering\arraybackslash}p{#1}}
\newcolumntype{L}[1]{>{\raggedright\arraybackslash}p{#1}}
\newcolumntype{R}[1]{>{\raggedleft\arraybackslash}p{#1}}
\newcolumntype{G}[1]{>{\columncolor{\altrowcolor}\centering\arraybackslash}p{#1}}

\newenvironment{task}{}{}
\newenvironment{category}{}{}

\newcommand{\codex}[1]{$\langle$#1$\rangle$}

\newcommand{\tnotex}{\textsuperscript}
\newcommand{\tablenotesx}[2]{\multicolumn{#1}{p{0.95\columnwidth}}{\tablenotesfontsize{#2}}}

\newcommand{\sig}[1]{\tnotex{\pnote}\makebox[-3pt][r]{\textbf{#1}}}
\newcommand{\ssig}[1]{\tnotex{\ppnote}\makebox[-5pt][r]{\textbf{#1}}}
\newcommand{\sssig}[1]{\tnotex{\pppnote}\makebox[-8pt][r]{\textbf{#1}}}

\begin{document}

\title[FIDO2 the Rescue?]{FIDO2 the Rescue? Platform vs. Roaming Authentication on Smartphones}

\author{Leon W{\"u}rsching}
\authornote{Both authors contributed equally to this research.}
\orcid{0000-0003-2648-6507}
\affiliation{%
    \institution{Technical University of Darmstadt}
    \streetaddress{Pankratiusstra{\ss}e 2}
    \city{Darmstadt}
    \country{Germany}
}
\email{lwuersching@seemoo.tu-darmstadt.de}

\author{Florentin Putz}
\orcid{0000-0003-3122-7315}
\authornotemark[1]
\affiliation{%
    \institution{Technical University of Darmstadt}
    \streetaddress{Pankratiusstra{\ss}e 2}
    \city{Darmstadt}
    \country{Germany}
}
\email{fputz@seemoo.tu-darmstadt.de}

\author{Steffen Haesler}
\orcid{0000-0002-6808-0487}
\affiliation{%
    \institution{Technical University of Darmstadt}
    \streetaddress{Pankratiusstra{\ss}e 2}
    \city{Darmstadt}
    \country{Germany}
}
\email{haesler@peasec.tu-darmstadt.de}

\author{Matthias Hollick}
\orcid{0000-0002-9163-5989}
\affiliation{%
    \institution{Technical University of Darmstadt}
    \streetaddress{Pankratiusstra{\ss}e 2}
    \city{Darmstadt}
    \country{Germany}
}
\email{mhollick@seemoo.tu-darmstadt.de}

\renewcommand{\shortauthors}{W{\"u}rsching and Putz, et al.}

\begin{abstract}
Modern smartphones support FIDO2 passwordless authentication using either external security keys or internal biometric authentication, but it is unclear whether users appreciate and accept these new forms of web authentication for their own accounts.
We present the first lab study (N=87) comparing platform and roaming authentication on smartphones, determining the practical strengths and weaknesses of FIDO2 as perceived by users in a mobile scenario.
Most participants were willing to adopt passwordless authentication during our in-person user study, but closer analysis shows that participants prioritize usability, security, and availability differently depending on the account type.
We identify remaining adoption barriers that prevent FIDO2 from succeeding password authentication, such as missing support for contemporary usage patterns, including account delegation and usage on multiple clients.
\end{abstract}


\keywords{Usability, Security, Passwordless, User Authentication, Biometrics, Accounts\\[3em]} %

\maketitle

\begin{figure}
	\begin{subfigure}[t]{0.49\columnwidth}
		\centering
		\includegraphics[width=\columnwidth]{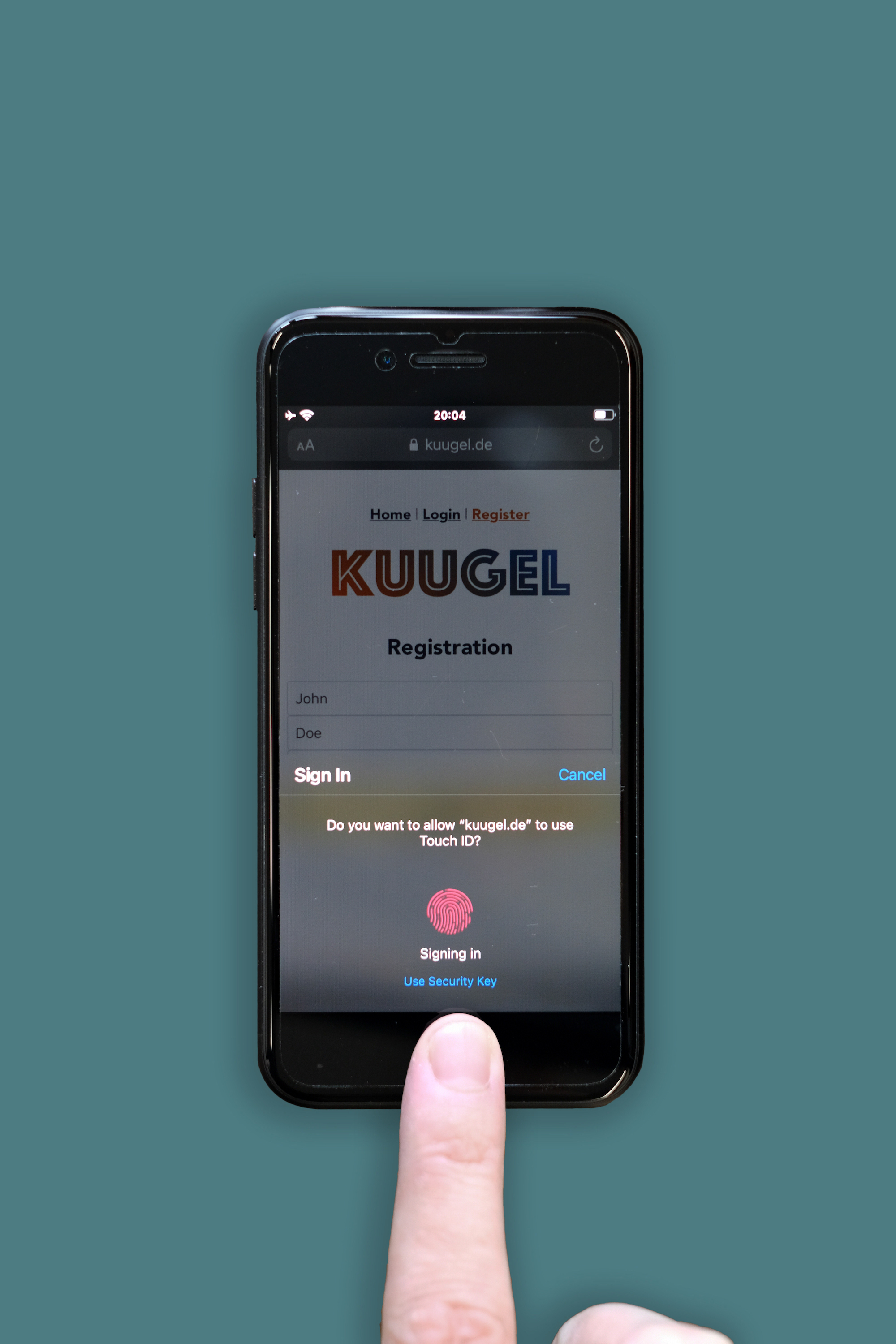}
		\caption{%
		    \groupP: Apple Touch ID
		}
		\Description{%
		    The Safari overlay GUI says "Do you want to allow "kuugel.de" to use Touch ID?" and shows a fingerprint.
		    On the bottom of the photo, a finger is lying on the iPhone's fingerprint sensor.
		}
		\label{fig:iphone-platform}
	\end{subfigure}
	\hfill
	\begin{subfigure}[t]{0.49\columnwidth}
		\centering
		\includegraphics[width=\columnwidth]{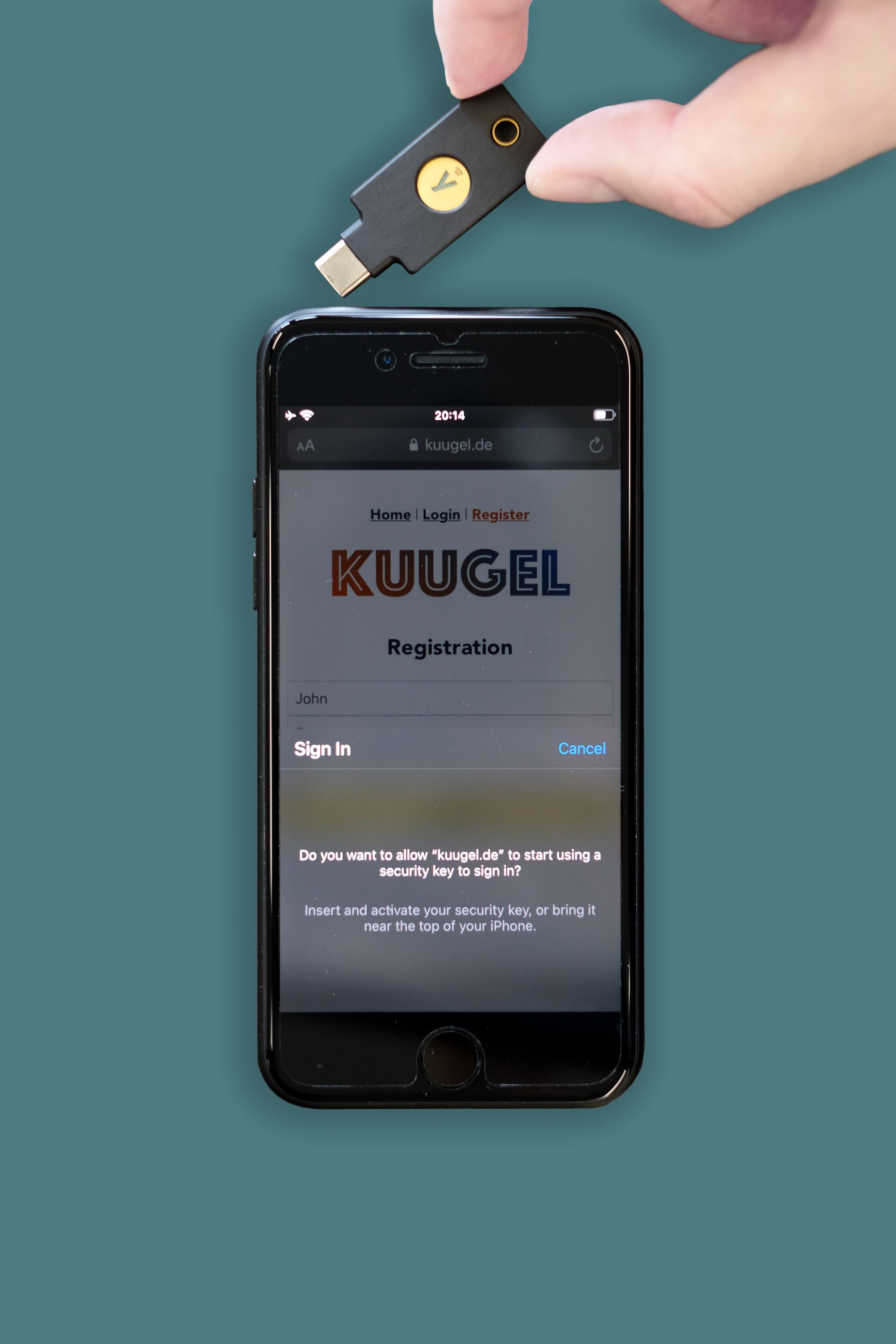}
		\caption{%
		    \groupR: Yubico Yubikey
	    }
		\Description{%
		    The Safari overlay GUI says "Do you want to allow "kuugel.de" to start using a security key to sign in? Insert and activate your security key, or bring it near the top of your iPhone."
		    On the top of the photo, two fingers are holding a YubiKey near the top of the iPhone.
		}
		\label{fig:iphone-roaming}
	\end{subfigure}
	\caption
	{%
	    The different user interactions with the smartphone during the experiment.
	    \groupP used Apple Touch ID as a platform authenticator and \groupR used a Yubico YubiKey as a roaming authenticator.
	}
	\Description{%
	    Two photos labelled (a) and (b) show how a user interacts with an iPhone during the registration on the KUUGEL website with platform and roaming authentication, respectively.
	    In the center of both photos, there is an Apple iPhone with the KUUGEL website opened in Safari.
	    On the bottom of the iPhone screen, both photos show an overlay Safari GUI with registration instructions.
	}
\end{figure}

\section{Introduction}
Authentication is one of the central building blocks for securing the Internet.
Since the 1970s, web services have conventionally relied on text-based passwords as the de-facto standard for user authentication.
Secure passwords, however, are hard to memorize \cite{yan2004password}, password reuse enables severe attacks \cite{ives2004domino}, and password-based authentication is prone to phishing \cite{dhamija2006phishing}.
Despite stronger security guarantees, all previous approaches in the \textquote{quest to replace passwords} failed to get widespread adoption, mainly because of their inferior deployability and usability \cite{bonneau2012quest}.

To address the first entry barrier \textendash~\emph{deployability}~\textendash~more than 250 technology companies and browser vendors \cite{webFidoMembers} have founded the \gls{FIDO}, jointly designing the FIDO2 standards for passwordless authentication \cite{webFido2}.
As a result, all major web browsers are now FIDO2-ready \cite{mdn2022webauthnCompatibility}, as they support the corresponding \acs{W3C} \gls{WebAuthn} standard \cite{standard2021webauthnLevel2}.
FIDO2 mandates public-key cryptography to provide user authentication based on authenticators containing the user's private key \cite{standard2021ctap}.
Unlike passwords, FIDO2 authentication is resistant to phishing, keylogging, replay attacks, and server breaches \cite{lang2016security}.

As more and more websites support FIDO2 authentication, it becomes essential to study the second entry barrier \textendash~\emph{usability}~\textendash~to find out how users react to this paradigm shift from knowledge-based factors to possession-based and biometrics-based factors.
There are two variants of FIDO2 authentication with fundamentally different user interactions:

For \emph{roaming authentication}, the private keys are stored on an external roaming authenticator, e.g., a YubiKey \cite{yubico2022yubikey5cNFC}, which connects to the client device via \gls{USB}, \gls{NFC}, or \gls{BLE}.
A recent usability study within the desktop environment by Lyastani et al. \cite{lyastani2020kingslayer} suggests that users accept and prefer roaming authentication as an alternative to passwords.
Many participants, however, were concerned about carrying a security key with them physically and potentially losing it.
Despite the prevalence of FIDO2 support in recent iPhones and Android devices \cite{fido20190225android,apple2020ios133},
\textbf{smartphones have not been studied as FIDO2 clients for roaming authentication yet}, but only as roaming authenticators themselves \cite{owens2021user,rasmussen2021usability}.

The second variant of FIDO2, \textit{platform authentication}, mitigates these availability concerns, as the smartphone's integrated \gls{TPM} stores the private keys, protected by an additional local authentication using the smartphone's unlock mechanism, e.g., Apple Touch ID \cite{apple2020touchidweb}.
However, platform authentication raises new usability concerns about the fundamentally different mental model.
Users need to grasp a more complex authentication method combining a private key stored on the smartphone's \gls{TPM} that is further protected by biometrics-based local authentication.
Platform authentication on smartphones has received little attention so far.
Oogami et al. \cite{oogami20observation} studied how to improve compatible websites' user experience, and Lassak et al. \cite{lassak2021mitigating} developed smartphone notifications addressing platform authentication misconceptions.
However, \textbf{the research community lacks an understanding of whether users understand, accept, and trust passwordless platform authentication as an alternative to roaming authentication and passwords}.

\begin{figure}
    \includegraphics[scale=.75]{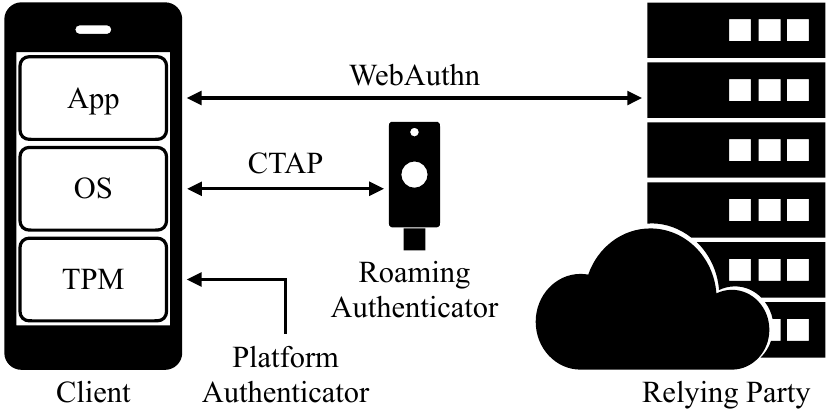}
    \caption
    {
    The parties and protocols involved in FIDO2 authentication scenarios.
    On the client side, \textquote{App} and \textquote{OS} are abbreviations for the application accessing the relying party and the client's operating system, respectively.
    }
    \Description{%
        Three components: Client, Authenticator, and Relying Party.
        The relying party is illustrated as a server and it is connected to the client via a bidirectional arrow labelled "WebAuthn".
        The client is illustrated as a smartphone and contains three labels for Application, Operating System, and Trusted Party Module.
        There is an arrow directed at the Trusted Party Module labelled "Platform Authenticator".
        A bidirectional arrow labelled "CTAP" connects the client to the roaming authenticator.
    }
    \label{fig:background-fido2-scenario}
\end{figure}

In this work, we try to bridge this knowledge gap by reporting, to the best of our knowledge, the first large-scale lab study comparing FIDO2 platform authentication and roaming authentication on smartphones.
We recruited 87 participants, randomly assigned them to one of two groups, and had them perform a series of practical web authentication tasks on an Apple iPhone.
The first group used platform authentication with Apple Touch ID for the web (\autoref{fig:iphone-platform}), while the second group used roaming authentication with an \gls{NFC}-based YubiKey (\autoref{fig:iphone-roaming}).
The participants reflected on their experience in a survey, which featured a combination of quantitative questions using standardized metrics and qualitative open-ended text questions.
This paper presents our work as follows:

\begin{itemize}[noitemsep]
    \item As our main contribution, we conduct the \textbf{first large-scale lab study that compares FIDO2 platform and roaming authentication on smartphones} (\autoref{sec:methods}).
    \item We show that platform and roaming authentication have excellent usability on smartphones, but lay users generally prefer platform authentication (\autoref{sec:results}).
    \item From our questionnaire's qualitative categories, we identify the strengths and weaknesses of FIDO2 on smartphones (\autoref{sec:results-qual}). Based on our participants' feedback, we investigate how to address the weaknesses of passwordless authentication and discuss account-specific adoption decisions and usage patterns (\autoref{sec:discussion}).
    \item We provide a replication package with our evaluation scripts and the pseudonymized dataset \cite{seemoo2023zenodo} collected in our study,
    consisting of 22 variables for each of our 87 participants.
    We also release our mockup website's source code \cite{seemoo2023github} to facilitate future work.
\end{itemize}

Our questionnaire encouraged participants to reflect on their everyday authentication use cases and whether they would be willing to use FIDO2 for their own accounts.
While most of our participants were generally willing to adopt passwordless authentication, account-specific adoption barriers remain for both roaming and platform authentication, which we discuss in detail.

\section{Background and Related Work}\label{sec:background-related-work} %

This section explains how the FIDO2 standards enable passwordless authentication on smartphones.
\autoref{ssec:background-fido2} describes relevant parts of the standards.
Then, we explain roaming authentication (\autoref{ssec:background-roaming}) and platform authentication (\autoref{ssec:background-platform}), including summaries of related studies of FIDO2.
We also consider authentication-related studies outside the FIDO2 cosmos (\autoref{ssec:rel-fido2-passwordless-misc}).

\subsection{FIDO2}\label{ssec:background-fido2}
The FIDO2 standards provide strong and passwordless authentication for the Internet.
All major web browsers support them \cite{mdn2022webauthnCompatibility}, and an increasing number of online services offer FIDO2 authentication both in \gls{1FA} and \gls{MFA} scenarios \cite{webFidoMembers}.
FIDO2 consists of two standards, \gls{WebAuthn} \cite{standard2021webauthnLevel2} and \gls{CTAP} \cite{standard2021ctap}, which specify the communication between relying party, the client, and the authenticator (\autoref{fig:background-fido2-scenario}).
Together they form a challenge-response protocol, confirming the user's identity to the relying party by verifying the user's possession of the authenticator that manages their public key credentials.

In our study, \gls{WebAuthn} handles the communication between a smartphone (client device) and a website (relying party).
However, FIDO2 is flexible and also supports local authentication scenarios, e.g., a user unlocking a Windows computer via Windows Hello \cite{fido2021windowshello}. 
On the client-side, \gls{WebAuthn} expects an authenticator with access to the secret key corresponding to the public key stored at the relying party.
FIDO2 supports two types of authenticators:

\subsection{Roaming Authentication}\label{ssec:background-roaming}
A roaming authenticator is a \gls{CTAP}-conforming external device that manages all public key credentials.
During authentication, the relying party's challenge is forwarded to the roaming authenticator, solved locally with the secret key, and returned to the relying party.
There is a large ecosystem of suitable roaming authenticators which are compatible with smartphones via \gls{NFC} \cite{yubico2022securitykeyNFC}, \gls{BLE} \cite{gotrust2022idemCard}, or Apple Lightning/\gls{USB} Type C \cite{feitian2022lightning}.
The smartphone itself must be equipped with a \gls{CTAP}-conforming interface and a \gls{WebAuthn}-compatible web browser.
Since 2019, FIDO2 roaming authentication has been supported on Android smartphones running Android 7 or later \cite{fido20190225android} and on Apple iPhones running iOS 13.3 or later via Apple Lightning \cite{yubico2020lightning} and \gls{NFC} \cite{apple2020ios133}.

\etal{Lyastani} \cite{lyastani2020kingslayer} conducted the first large-scale user study
of FIDO2 roaming authentication on computers.
In their between-groups study, participants either authenticated with passwords or used a YubiKey, concluding that roaming authentication is more usable and accepted than passwords.
\etal{Farke} \cite{farke2020afterAll} studied FIDO2 roaming authentication as an unlocking mechanism for computers, using YubiKeys with enabled \gls{PIN} protection.
They reported that participants stopped using roaming authentication because it was slower than their password manager.

Previous studies investigated whether smartphones, themselves, could be used as external FIDO2 roaming authenticators to address deployment and availability issues of traditional security keys.
Owens et al. \cite{owens2020smartphones,owens2021user} reported a between-groups observation study,
comparing passwords to smartphones as roaming authenticators.
They concluded that users understand the security benefits of FIDO2 but still find password-based authentication more usable.
Rasmussen \cite{rasmussen2021usability} conducted a between-groups user study
showing that smartphones as roaming authenticators have similar usability and acceptance as YubiKeys.
Both studies \cite{owens2021user,rasmussen2021usability} reported availability concerns regarding empty smartphone batteries.

We continue with a brief overview of user studies on the \gls{U2F} \cite{standard2017u2f}, which is \gls{CTAP}'s predecessor and has similar user interaction.
\etal{Lang} \cite{lang2016security} reported on a two-year enterprise deployment of security keys within Google, laying the foundation for the \gls{U2F} standard
that is \gls{CTAP}'s predecessor.
\etal{Das} \cite{das2018qualitative,das2018johnny} studied \gls{U2F} in non-enterprise environments and found that a \gls{2FA} method's acceptance did not correlate with its usability.
\etal{Reynolds} \cite{reynolds2018tale} and \etal{Reese} \cite{reese2019usability} conducted longitudinal studies of \gls{U2F}, reporting that the initial setup is cumbersome compared to the fast and easy authentication afterward.
\etal{Ciolino} and \etal{Das} investigated lay users' perception of \gls{U2F},
finding that, while sentiment towards \gls{U2F} is lower than for \acs{SMS}-based \gls{2FA} \cite{ciolino2019twoMinds}, some lay users do not feel the need to protect their accounts with \gls{MFA} at all \cite{das2019towards}.
\etal{Colnago} \cite{colnago2018horrible} conducted a large-scale longitudinal study to explore \gls{2FA} adoption at a university, finding that $<1\%$ use \gls{U2F} as a \gls{2FA} method.
More recent works on \gls{U2F} by \etal{Das}  and \etal{Reynolds}
found that users are more likely to adopt \gls{MFA} for essential accounts \cite{das2020necessaryChore} and that the users' initial negative perception of \gls{MFA} methods fades over time \cite{reynolds2020empirical}.

\subsection{Platform Authentication}\label{ssec:background-platform}
Client devices qualified to manage cryptographic key pairs can be used as platform authenticators.
During authentication, the relying party's challenge is solved directly on the client's platform authenticator.
Most platform authenticators require the user to authenticate locally with a \gls{PIN} or biometrics, augmenting FIDO2 to an \gls{MFA} method.
Any Android phone running Android 7 or later provides a FIDO2 platform authenticator, as Android's FIDO2 certification includes the built-in biometrics sensors \cite{fido20190225android}.
Since iOS 14, Apple has equipped iPhones with a platform authenticator, namely Touch ID for the web \cite{apple2020ios14}, which we also refer to as \emph{Touch ID} for simplicity.
For Touch ID platform authentication, the iPhone's Secure Enclave manages FIDO2 credentials and requires the user to authenticate locally with Touch ID.

\etal{Oogami} \cite{oogami20observation} studied the usability of FIDO2 platform authentication on smartphones, conducting interviews to improve the website user experience for first-time FIDO2 users.
However, they did not collect FIDO2 weaknesses or adoption barriers of platform authentication.
\etal{Lassak} \cite{lassak2021mitigating} studied misconceptions about FIDO2 platform authentication on Android phones and developed smartphone notifications to address them.

\subsection{Miscellaneous}\label{ssec:rel-fido2-passwordless-misc}
Independently of FIDO2, van den Boogaard \cite{boogaard2022user} conducted an online study
of users' understanding of biometrics-based authentication on mobile phones, finding that most participants use biometric options when available.
\etal{Conners} \cite{conners2022letsAuthenticate} conducted an online study
of \emph{Let's authenticate} \cite{conners2019letsAuthenticate}, which is similar to FIDO2 but uses authenticators (roaming or platform) to issue certificates for user authentication.
\etal{Alqubaisi} \cite{alqubaisi2020RushImplement} studied elements hindering the adoption of FIDO2 passwordless authentication, basing their analysis on the \gls{FIDO} developer mailing list.

\section{Research Questions}\label{sec:research-questions}
We continue the work of recent FIDO2 user studies but focus on passwordless authentication on smartphone clients.
Thus, the research questions addressed in this work are:

\begin{itemize}
\item\textbf{RQ1}
\emph{What is the usability and acceptance of FIDO2 passwordless authentication using platform and roaming authentication on smartphones?}

\item\textbf{RQ2}
\emph{What benefits and concerns do users consider when using FIDO2 passwordless authentication on smartphones?}

\item\textbf{RQ3}
\emph{Which account types do users want to secure using FIDO2 passwordless authentication on smartphones?}
\end{itemize}
In contrast to related work, to the best of our knowledge, we are the first to study FIDO2 roaming authentication on smartphone clients and the first to conduct a lab study on any device type that compares FIDO2 platform authentication and roaming authentication.
Our main goal is to identify the strengths and weaknesses of passwordless authentication on smartphones. To this end, we determine account-specific adoption barriers by encouraging our participants to reflect on whether they want to use passwordless authentication for their own accounts.

\begin{figure}
\centering
	\begin{subfigure}[t]{0.49\columnwidth}
		\centering
		\includegraphics[width=\columnwidth]{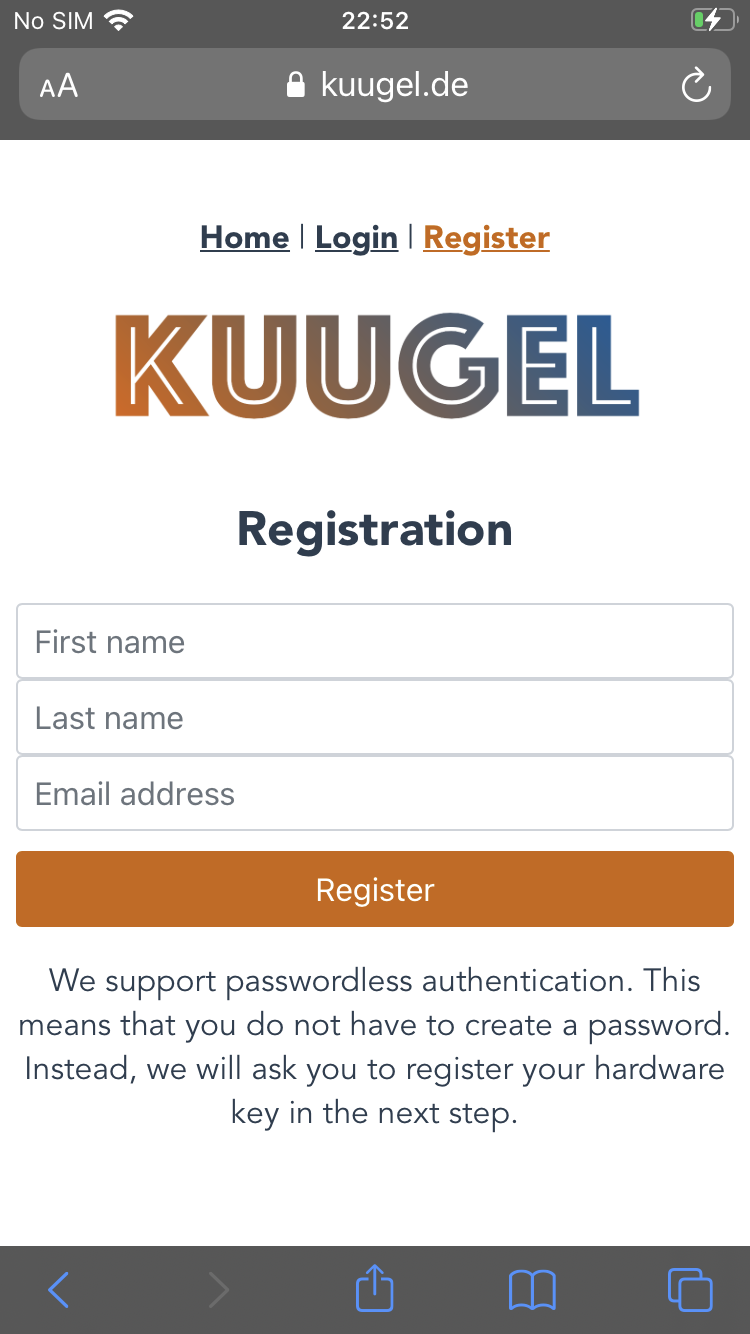}
		\caption{Registration page.}
		\Description{%
		    The registration form includes three fields: first name, last name, email address.
		    The button says "register" and the follow-up text explains "We support passwordless authentication. This means that you do not have to create a password. Instead, we will ask you to register your hardware key in the next step".
		}
		\label{fig:kuugel-register}
	\end{subfigure}
	\hfill
	\begin{subfigure}[t]{0.49\columnwidth}
		\centering
		\includegraphics[width=\columnwidth]{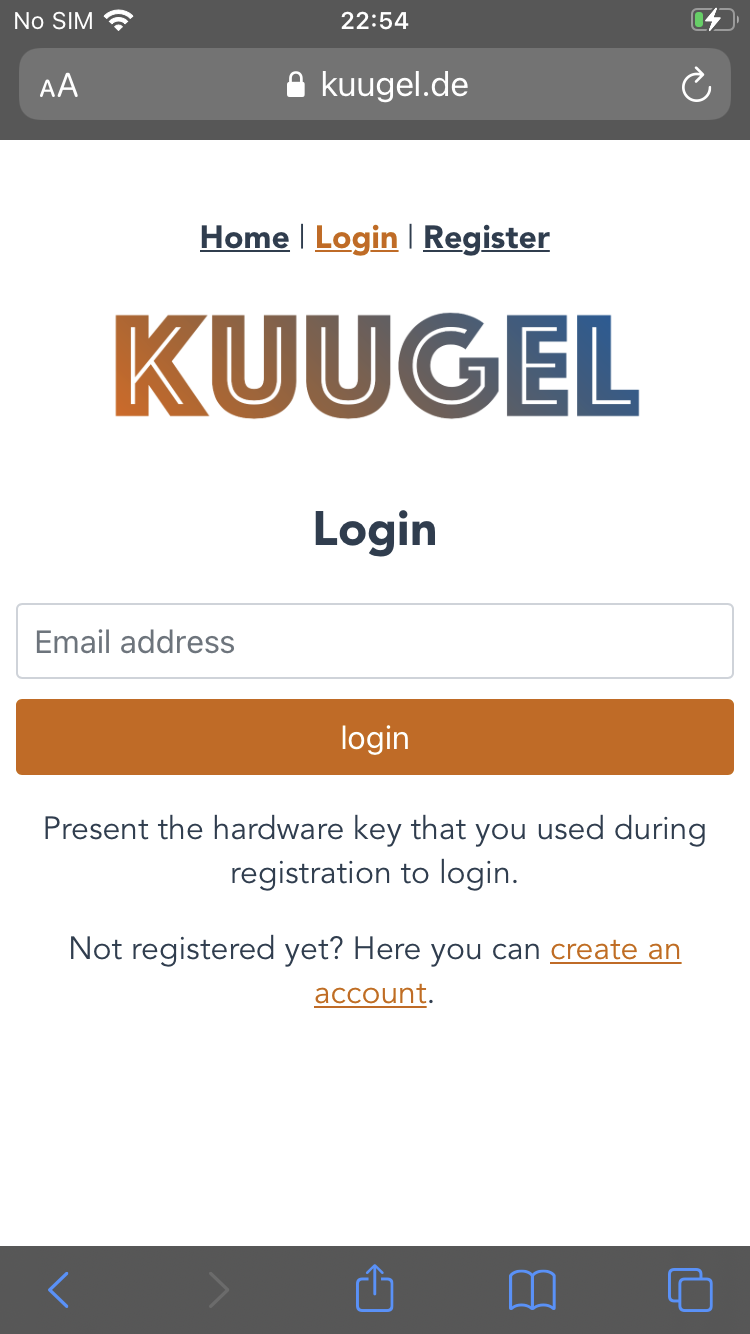}
		\caption{Login page.}
		\Description{%
		    The login form includes a single field: "Email address".
		    The button says "login" and the follow-up text explains "Present the hardware key that you used during registration to login. Not registered yet? Here you can create an account." where the last three words "create an account" are a hyperlink.
		}
		\label{fig:kuugel-login}
	\end{subfigure}
	\caption
	[Registration and Login of our Relying Party]
	{Screenshots of our mockup website's registration and login page. In our experiment, each participant experienced the complete authentication flow of FIDO2 passwordless authentication consisting of account registration and subsequent login.}
		\Description{%
		    Two screenshots from an iPhone with Safari open.
		    The screenshots (a) and (b) are showing the KUUGEL website's registration and login page, respectively.
		    On both screenshots, there is a form and button in the center of the screenshot, followed by explanatory text.
		}
	\label{fig:kuugel}
\end{figure}

\section{Methods}\label{sec:methods}

To answer our research questions on the usability and acceptance of platform and roaming authentication on smartphones, we conducted a lab study with practical web authentication tasks (\autoref{ssec:methods-material}) and a follow-up survey containing quantitative and qualitative questions (\autoref{ssec:study-design}).
We address our study sample (\autoref{ssec:methods-participants}) and ethical considerations (\autoref{ssec:methods-ethics}), including the pilot study (\autoref{ssec:pilot-study}), and analysis toolbox (\autoref{sec:methods-analysis}).

\subsection{Material}\label{ssec:methods-material}

Our lab study's goal was to create a passwordless authentication scenario resembling our participants' everyday lives.
In our study, the participants gained hands-on authentication experience on a smartphone, namely the Apple iPhone SE (2nd generation), referred to as \emph{iPhone} for simplicity, running iOS 14.5.1 and Safari 14.1.
During our study, this setup represented the most popular mobile operating system in North America \cite{statista2022shareMobileOS}. 
We developed the mockup website \kuugel as a relying party (\autoref{fig:kuugel}), which supports FIDO2 registration/login, e.g., using either (1) platform authentication with Touch ID \cite{apple2020touchidweb} or (2) roaming authentication with a Yubico YubiKey 5C NFC \cite{yubico2022yubikey5cNFC}, referred to as \emph{YubiKey} for simplicity.
The YubiKey supports FIDO2 via an \gls{NFC} interface.
Furthermore, it features optional \gls{PIN} protection, adding a knowledge-based authentication factor.
Farke et al. \cite{farke2020afterAll} reported negative user feedback on this feature.
Thus, we opted against activating the YubiKey's \gls{PIN} protection to avoid additional authentication overheads.

\begin{figure*}[t]
	\includegraphics[scale=.9]{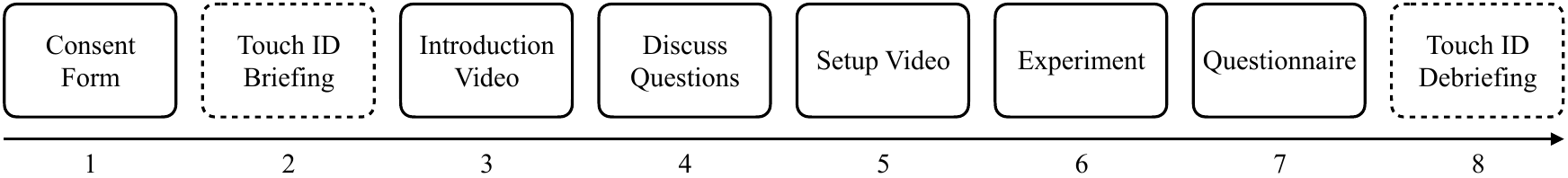}
	\caption{%
    Our lab study's eight stages.
    Dashed borders indicate stages that only participants of \groupP participated in to register and delete their fingerprints from the iPhone's Touch ID settings.
	}
		\Description{%
		    A flow graph depicting eight boxes from left to right: (1) Consent Form, (2) Touch ID Briefing, (3) Introduction Video, (4) Discuss Questions, (5) Setup Video, (6) Experiment, (7) Questionnaire, and (8) Touch ID Debriefing.
		    The borders of the boxes (2) Touch ID Briefing and  (8) Touch ID Debriefing are dashed, indicating that only participants of \groupP participated in these stages.
		}
	\label{fig:study-stages}
\end{figure*}

\subsection{Study Design}\label{ssec:study-design}
We used a between-groups design to identify the differences between platform and roaming authentication on smartphones.
Our between-groups design eliminates the influence of order and concentration level with an extended study duration.
We decided for a lab study over alternative study designs to ensure consistent conditions for all participants, minimizing confounding factors.
We randomly distributed the participants into two groups:
The study group used platform authentication with Touch ID (\groupP), while the control group used roaming authentication with a YubiKey (\groupR).
We conducted our lab study in neutral meeting rooms, where the participants met the study conductor in one-on-one sessions\footnote{Protocols were in place for the in-person lab study ensuring the safety of participants and the study conductor, as we conducted our study during the COVID-19 pandemic.}.
Each room contained a chair in front of a desk with an iPhone and, for \groupR, also a YubiKey.
For all participants, we ensured the same neutral environment and hardware to help participants only evaluate the effect of the presented authentication method.
The study conductor was always present during each trial to verify that the participant performed the tasks and watched the educational videos.

Our study consisted of eight stages, which are shown in \autoref{fig:study-stages}.
We first asked participants to read and sign our consent form, which explained the study's goal, procedure, and privacy policy.
We then instructed participants in \groupP to register one of their fingerprints in the iPhone's Touch ID settings, which was required for the following experiment.
We explicitly decided to do this step early before the introductory videos and practical tasks so that the rest of the study procedure was comparable for both groups.

\subsubsection{Task Instruction}
All participants watched group-specific educational videos (approx.~5 minutes total) before the experiment, introducing FIDO2 passwordless authentication and ensuring that all participants had identical conditions\footnote{Our introduction video was partly based on the introductory videos from the study of Lyastani et al. \cite{lyastani2020kingslayer}.}.
The video continued with a brief explanation of our mockup website's registration and login process.
We decided to use pre-recorded videos for instruction to increase the internal validity of our study by ensuring that all participants received the same information.
Both groups watched identically structured videos that only differed in the platform and roaming authentication details, allowing for fair comparison in our between-groups design.
Our goal was to avoid misunderstandings resulting in confounding effects, that are invisible in the worst case.

\subsubsection{Experiment}
After the task instruction, we handed our participants the smartphone (and for \groupR, also the YubiKey).
The participants used the iPhone to browse our mockup website and register an account using their assigned authenticator type.
Our mockup website complies with Apple's recommendations for websites with Touch ID for the Web \cite{apple2020touchidweb}.
Besides the pages necessary for registration and login, the mockup website provides only minimal additional functionality to help participants assess only the usability of FIDO2.

After registration, participants in \groupP logged into our mockup website with an email address and their fingerprint, as shown in \autoref{fig:iphone-platform}.
Participants in \groupR used their email address and the YubiKey for authentication at our mockup website.
The user brings the YubiKey near the iPhone to register a new credential on the YubiKey, as shown in \autoref{fig:iphone-roaming}.
After registration, the participants logged into their accounts using the same method.

\subsubsection{Questionnaire}
After collecting the hardware, we asked the participants to fill in our questionnaire to reflect on their experiences during the experiment.
\autoref{sec:questionnaire} contains the corresponding questionnaire.
We included variables from related work on authentication factors \cite{lyastani2020kingslayer, oogami20observation, reynolds2018tale, rasmussen2021usability} for comparability with previous user studies:

\begin{itemize}
\descitem{Dependent Variables}
The dependent variables captured our participants' experience and assessment of roaming or platform authentication.
To answer RQ1, we used the \gls{SUS} \cite{brooke1996sus} to determine the authentication methods' usability, as well as the acceptance scale from van der Laan et al. \cite{van1997simple} to measure how much our participants accepted their assigned authentication method.
To answer RQ3, we used 11 five-level Likert items to measure how likely our participants would use the demonstrated authentication method on 11 different types of online accounts.
As support for FIDO2 is currently limited in practice, we asked our participants to assume that each service supported the presented authentication method.
Participants could also select "not available" if they did not use this type of account.

\descitem{General Impression}
Our questionnaire included four open-ended text questions to answer RQ2. Participants reflected on their general impression of the studied authentication method, but we also specifically asked our participants to state any strengths and weaknesses that came to their minds. Our fourth text question, which directly follows the quantitative question on adoption for 11 different account types, was designed to learn why participants do (not) want to use the authentication methods for specific account types.
Our goal was to recognize account characteristics increasing or decreasing participants' likelihood of using passwordless authentication and identify adoption barriers for common account types.

\descitem{Control Variables}
As control variables, we measured our participants' technology affinity using the \gls{ATI} \cite{franke2019affinity} as well as their level of privacy concerns using four Likert items from Langer et al. \cite{langer2018information}.
Furthermore, we asked participants which \gls{2FA} methods they had already used and whether they had prior experience with Apple iOS, which was the operating system running on our study's smartphone.

\descitem{Demographic Variables}
We collected the gender, age, education, and field of study/work of our participants as demographic variables.

\end{itemize}

After filling in the questionnaire, we ensured participants of \groupP deleted their fingerprints from the iPhone's Touch ID settings.
Before concluding the study, we thanked the participants for their time.
It usually took participants 15-25 minutes to participate in our study.

\subsection{Participants}\label{ssec:methods-participants}
We conducted our lab study with 89 participants between July and September 2021.
For recruitment, we used mailing lists, social media groups, word-of-mouth, and snowball sampling, both within and outside our university (participants had to be over 18 years old to be eligible).
Interested participants self-registered with a registration form to choose a time slot for voluntary participation in our study.
Participants did not receive compensation for their participation in our study, except for one student from our university's psychology department who received a participation point as part of their study program.
Within this study program, the student had sufficient studies to choose from.

Two participants did not fully complete the questionnaire, so we removed them from our final sample (N = 87).
Of these, 55 (63\%) identified as male and 32 (37\%) as female, with no participants opting for the \textquote{other} or \textquote{no answer} options.
The education level of our participants was high, with the most common highest educational degree in our sample being a Bachelor's degree (40\%; 35).
Our diverse recruitment sources resulted in at least 39 (45\%) of non-students.
Most participants (59\%; 51) were between 20 and 29 years old, but a substantial share of participants (34\%; 30) were older than 30, of which 10 participants (11\%) were over 50.
According to Pearson's chi-squared test, the demographic structure (gender, age, and education) did not differ significantly between \groupP and \groupR (\autoref{tab:demographics}).

\subsection{Ethical Concerns}\label{ssec:methods-ethics}
Our university's \gls{IRB} reviewed and approved this study.
We informed all participants about the study's purpose and data collection while adhering to the \gls{GDPR}. We collected written consent prior to the lab study.

During the study, participants in \groupP temporarily registered their fingerprints on the iPhone. We briefed the participants about this and ensured the deletion of their fingerprints after each trial. For participants in \groupR, we reset the YubiKey to factory settings after each trial to ensure equal conditions for all participants.
The names and email addresses for authentication at our mockup website were only temporarily stored on the iPhone and deleted after each trial.
We did not collect any other sensitive information.
We pseudonymously stored each participant's answers without identifying information by assigning sequential numbers.
Participation in our study was voluntary and without compensation in accordance with common practice of our university's computer science department.

\subsection{Pilot Study}\label{ssec:pilot-study}
We conducted a pilot study (N=6) to identify technical problems with the study setup and ambiguities in the instructional videos and the questionnaire.
The pilot study showed that the setup worked reliably and that the participants understood the instructional videos and all categories of the questionnaire.

\begin{figure}
\begin{subfigure}[t]{\columnwidth}
    \centering
		\includegraphics[scale=0.75]{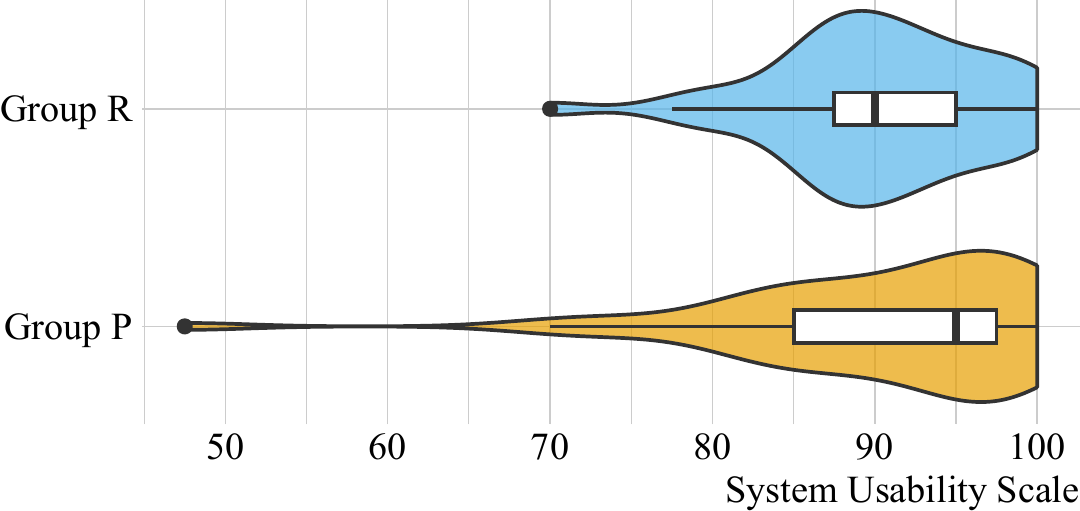}
    \caption
    {
    \gls{SUS} scores.
    }
    \Description{%
        Two violin plots showing the density estimation of the distribution of our participants' SUS scores for \groupR and \groupP, respectively.
        Median values and the interquartile ranges are provided in \autoref{tab:descriptive}.
        \groupP has a higher median SUS score.
        For \groupR, there is one outlier at a SUS of 70.
        For \groupP, there is one outlier at a SUS of 47.5.
	}
    \label{fig:evaluation-sus}
\end{subfigure}
\begin{subfigure}[t]{\columnwidth}
    \centering
		\includegraphics[scale=.75]{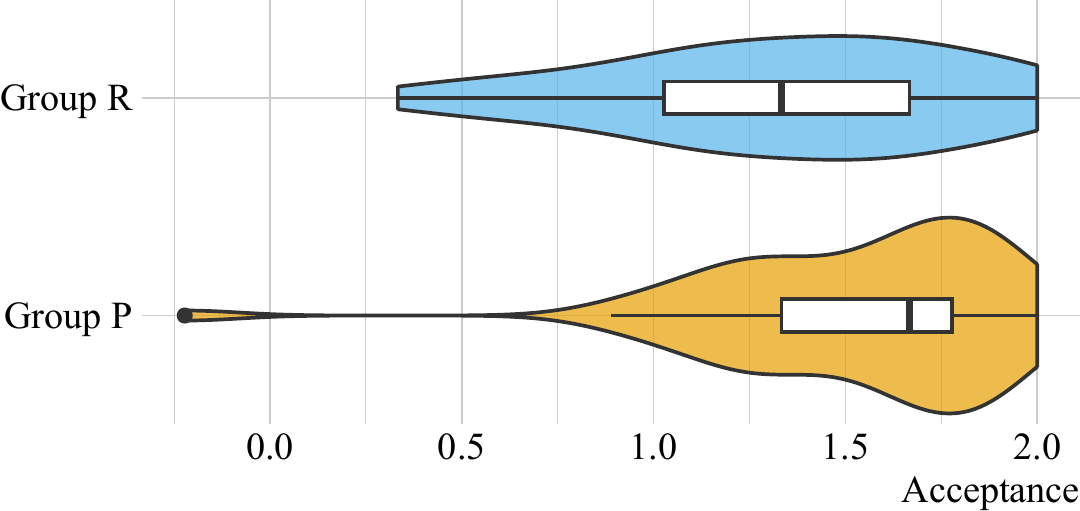}
    \caption
    {
    Acceptance scores.
    }
    \Description{%
        Two violin plots showing the density estimation of the distribution of our participants' acceptance scores for \groupP and \groupR, respectively.
        Median values and the interquartile ranges are provided in \autoref{tab:descriptive}.
        \groupP has a higher median acceptance score and one outlier at -0.22.
        \groupR has no outliers.
	}
    \label{fig:evaluation-acc}
\end{subfigure}
\caption%
{
Comparison of our participants' \gls{SUS} scores and acceptance scores for platform authentication (\groupP) and roaming authentication (\groupR).
The boxplots show quartiles, median, and outliers.
}
\Description{%
    Two figures labelled (a) and (b) showing the distribution of SUS and acceptance scores, respectively.
}
\end{figure}

\subsection{Data Analysis}\label{sec:methods-analysis}
We used the statistical software R 4.2.0 for all quantitative data analysis \cite{r2022language}.
We calculated the central tendencies and correlations to answer our research questions.
We compared \groupP and \groupR using the Wilcoxon-Mann-Whitney test \cite{mann1947test} for ordinal variables and Pearson's chi-squared test \cite{pearson1900criterion} for nominal variables.
For multiple tests, we controlled the \gls{FDR} using the Benjamini-Hochberg method \cite{benjamini1995controlling}.
We used Kendall's rank correlation coefficient \cite{kendall1938measure} to determine the relationship between our control and dependent variables.
For all statistical tests, we used an alpha level of .05.

Furthermore, we conducted a qualitative analysis of the four open-ended text questions,
performing the following qualitative coding steps:
(1) Two researchers independently constructed an initial codebook for each question using inductive coding \cite{gioia2013seeking,merriam2015qualitative}.
(2) The researchers merged these codebooks through discussions and formed clusters to identify a suitable level of detail. (3) Two further independent researchers deductively coded all data according to the final codebook. (4) They discussed to resolve coding differences.
After discussions, the researchers agreed on most coding decisions and reached satisfactory inter-coder reliability (mean Krippendorff's Alpha = .984, minimum .789 \cite{krippendorff2004content}). Finally, the researchers who created the initial codebook discussed and agreed on any remaining coding inconsistencies.
\autoref{sec:codebooks} contains our final coding system.

\section{Quantitative Results}\label{sec:results}

In this section, we report on the evaluation of our questionnaire, which represents the results of our study.
The questionnaire (\autoref{ssec:study-design}) contains quantitative scales, control variables, and open-ended qualitative questions.
We used well-known metrics \cite{brooke1996sus,van1997simple} to study the participants' perceptions of FIDO2 regarding its usability (\autoref{ssec:quan-usability}) and acceptance (\autoref{ssec:quan-acceptance}).
Furthermore, we determined the participants' likelihood to adopt FIDO2 for specific online account types (\autoref{ssec:quan-adoption} and analyzed the control variables (\autoref{sec:control-variables}).

\subsection{Usability}\label{ssec:quan-usability}

After getting hands-on experience with passwordless authentication on smartphones, the participants completed our questionnaire, including the \gls{SUS} \cite{brooke1996sus}.
We use the Shapiro-Wilk test \cite{shapiro1965analysis} to determine whether this variable is normally distributed.
The \gls{SUS} scores for \groupR ($W=0.95$, $p=.058$) are approximately normal, but the scores for \groupP ($W=0.81$, $p<.001$) are significantly non-normal.
\autoref{fig:evaluation-sus} depicts the statistics of the \gls{SUS} scores per group. The distribution in \groupP has visible negative skew, indicating that most participants selected values on the higher end of the scale. Additionally, there is a single outlier in both groups.
Parametric methods such as the t-test assume normally distributed sampling distributions and can give inaccurate results in the presence of outliers \cite{wilcox2017robust}. We, therefore, use non-parametric statistical methods for data analysis (\autoref{sec:methods-analysis}).

\gls{SUS} scores in \groupP (Mdn = 95) did not differ significantly from \groupR (Mdn = 90) according to the Wilcoxon-Mann-Whitney test, $W = 835.5$, $p=.351$, $r=-.10$.
\textbf{Both platform and roaming authentication on smartphones have a similarly high level of usability.}

\subsection{Acceptance}\label{ssec:quan-acceptance}
We collected data on the acceptance of passwordless authentication using the scale of \etal{van der Laan} \cite{van1997simple}.
\autoref{fig:evaluation-acc} shows the acceptance scores for both groups, which are approximately normal for \groupR ($W=0.96$, $p=.196$), but significantly non-normal for \groupP ($W=0.83$, $p<.001$). As with the \gls{SUS} scores, the acceptance scores in \groupP have strong negative skew and an outlier, which reaffirms our choice of the non-parametric Wilcoxon-Mann-Whitney test.
The acceptance scores in \groupP (Mdn = 1.7) differed significantly from \groupR (Mdn = 1.3), $W = 691$, $p=.031$, $r=-.23$, indicating that \textbf{platform authentication has higher acceptance on smartphones than roaming authentication}.

\begin{figure*}[t]
    \centering
		\includegraphics[scale=0.75]{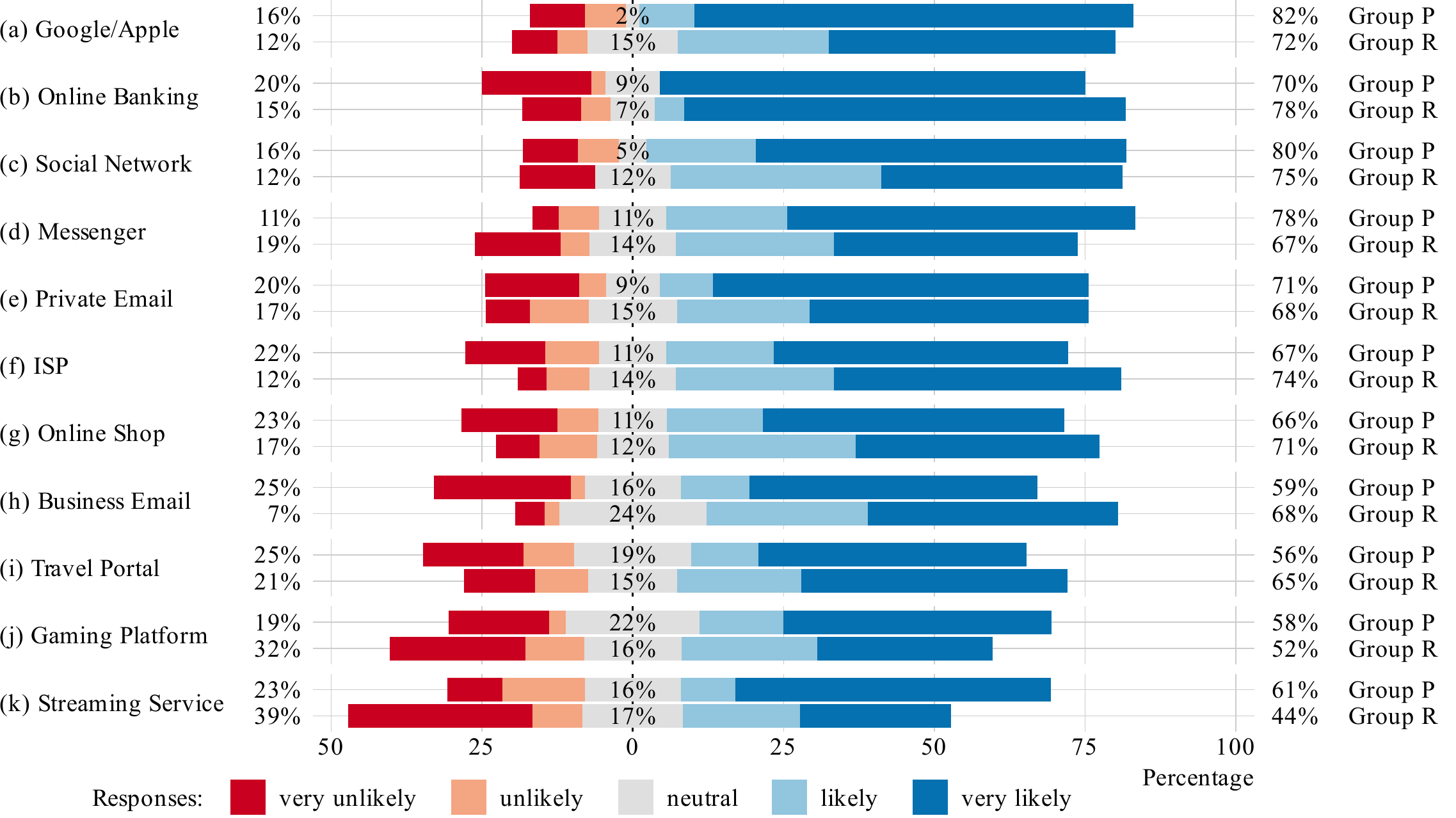}
    \caption[Account Adoption Scores]
    {
    Comparison of our participants' adoption likelihood for different account types ordered by their average likelihood.
    The percentages (left, middle, right) represent the share of negative, neutral, and positive responses, respectively.
    }
    \Description{%
        Diverged bar plots showing our participants' adoption likelihood for 11 different account types.
        These accounts are ordered by their average adoption likelihood with the largest first, (a) to (k).
        For each account type and study group, we list the share of negative, neutral, and positive responses.
        (a) Google/Apple (P 16\%, 2\%, 82\%; R 12\%, 15\%, 72\%),
        (b) Online Banking (P 20\%, 9\%, 70\%; R 15\%, 7\%, 78\%),
        (c) Social Network (P 16\%, 5\%, 80\%; R 15\%, 7\%, 78\%),
        (d) Messenger (P 11\%, 11\%, 78\%; R 19\%, 14\%, 67\%),
        (e) Private Email (P 20\%, 9\%, 71\%; R 17\%, 15\%, 68\%),
        (f) ISP (P 22\%, 11\%, 67\%; R 12\%, 14\%, 74\%),
        (g) Online Shop (P 23\%, 11\%, 66\%; R 17\%, 12\%, 71\%),
        (h) Business Email (P 25\%, 16\%, 69\%; R 7\%, 24\%, 68\%),
        (i) Travel Portal (P 25\%, 19\%, 56\%; R 21\%, 15\%, 65\%),
        (j) Gaming Platform (P 19\%, 22\%, 58\%; R 32\%, 16\%, 52\%),
        (k) Streaming Service (P 23\%, 16\%, 61\%; R 39\%, 17\%, 44\%).
    }
    \label{fig:evaluation-adoption-likert}
\end{figure*}

\subsection{Adoption}\label{ssec:quan-adoption}

The category dedicated to adoption listed 11 account types and asked participants how likely they were to use the studied passwordless authentication method, provided that the online service supported it.
\autoref{fig:evaluation-adoption-likert} shows the responses for all accounts.

Most participants were optimistic about using passwordless authentication on their smartphones.
The three account types with the highest share of positive responses were Google/Apple accounts (77\%; 65), online banking accounts (74\%; 63), and social network accounts (77\%; 65), although the last had less \textquote{very likely} responses compared to the former two.
Our participants were least likely to use passwordless authentication for streaming services (54\%; 43 positive responses) and gaming platforms (55\%; 37 positive responses).
For every account type, however, the share of positive responses was higher than the share of negative responses, indicating that \textbf{overall, our participants want to use passwordless authentication on smartphones when possible.}

We compared the adoption scores of both groups using Pearson's chi-squared test. The adoption did not differ significantly between both groups for any account type
after controlling the \gls{FDR} using the Benjamini-Hochberg method and a target \gls{FDR} of 10\% \cite{benjamini1995controlling}.
We excluded participants without a certain account from the analysis of that account type.
This occurred only for a few participants per account, with the following exceptions: Twenty participants did not have a gaming account (11 P + 9 R), 17 participants did not have an account at a travel portal (9 P + 8 R), and 7 participants did not have an account at a streaming service (1 P + 6 R).

\subsection{Control Variables}
\label{sec:control-variables}
Our control variables are part of the descriptive overview in \autoref{tab:descriptive}.
Additionally, \autoref{sec:demographics} contains our participants' demographic information.
\groupP and \groupR did not differ significantly for \gls{ATI}, privacy concerns, and iOS familiarity.
We calculate bivariate Kendall's rank correlation coefficients in \autoref{tab:correlations} to determine the relationship between the control and dependent variables.
As shown, the acceptance scores significantly correlated with the \gls{SUS} scores ($\tau = .36, p<.001$) and the predictor variable representing the differences between the two groups ($\tau = .20, p=.030$). There also was a significant negative correlation between the \gls{ATI} scores and iPhone familiarity ($\tau=-.23, p=.009$).

\briefSection{2FA Familiarity}
We asked our participants whether they were familiar with any \gls{2FA} methods. Most of our participants had prior experience with \acs{SMS}-based \gls{2FA} (87\%; 76), push-based smartphone apps (83\%; 72), and \gls{TAN} lists (76\%; 66). About half had used \gls{OTP} generator apps before (51\%; 44), and only a few had experience with security keys such as the YubiKey (8\%; 7).
\groupP and \groupR did not differ significantly for any of the five \gls{2FA} methods.

\begin{table}
\centering
\caption{Comparison of our participants' descriptive data, including the control and dependent variables.}
\tablefontsize
\begin{tabularx}{\columnwidth}{Xrrrr}
\toprule
\tableheadline{Variable} &
\tableheadline{\groupP} &
\tableheadline{\groupR} &
\tableheadline{Statistic} &
\tableheadline{ES}\\
& ($N=45$) & ($N=42$) \\
\midrule
iOS Familiarity &&&$\chi^2(1)=0.05$&.03\\
\qquad Yes & 67\% (30) & 64\% (27) & $p=.815$&\\
\qquad No & 33\% (15) & 36\% (15) &&\\
\rowcolor{\altrowcolor} ATI & 4.2 & 4.5 &$W = 1072$&-.12\\
\rowcolor{\altrowcolor}     & (2.2) & (1.5) & $p=.282$&\\
Privacy Concerns & 5.0 & 5.0 &$W = 966$&-.02\\
    & (1.8) & (1.8) & $p=.861$&\\
 \midrule
\rowcolor{\altrowcolor} SUS & 95 & 90 &$W = 835.5$&-.10\\
\rowcolor{\altrowcolor}     & (12.5) & (7.5) & $p=.351$&\\
Acceptance & 1.7 & 1.3 &$W = 691$&-.23\\
    & (0.4) & (0.6) & $\bm{p=.031}$&\\
\bottomrule
\tablenotesx{5}{\textit{Note:} For iOS familiarity, we report in-group percentages (and frequencies), Pearson's chi-square test, and Cram\'er's V \cite{cramer1946mathematical} as the \gls{ES}. For the other scores based on Likert scales, we report the median (and the interquartile range), the Wilcoxon-Mann-Whitney test, and the effect size estimate based on Rosenthal's method \cite{rosenthal1991meta}.}
\end{tabularx}
\label{tab:descriptive}
\end{table}
\begin{table}
\centering
\caption[Correlations]
{
  Kendall's correlation between control and dependent variables.
}
\tablefontsize
\setlength{\tabcolsep}{8.5pt}
\begin{tabularx}{\columnwidth}{@{} r@{\hspace{1\tabcolsep}} X @{\hspace{0.5\tabcolsep}}rrr @{\hspace{2\tabcolsep}} rrr @{}}
\toprule
&&
\multicolumn{1}{c}{\tableheadline{2}} &
\multicolumn{1}{c}{\tableheadline{3}} &
\multicolumn{1}{c}{\tableheadline{4}} &
\multicolumn{1}{c}{\tableheadline{5}} &
\multicolumn{2}{c}{\tableheadline{Accept.}} \\
\midrule
1 & Group (P) &  -.10 & -.02 &  .03 &  .09 &  \sig{.20} & \\
\rowcolor{\altrowcolor} 2 & ATI       &       &  .05 & \ssig{-.23} &  .15 &  .12 & \\
3 & Privacy Concerns  &       &      & -.09 & -.12 & -.14 & \\
\rowcolor{\altrowcolor} 4 & iOS Familiarity    &       &      &      &  .00 & -.02 & \\
5 & SUS       &       &      &      &      & \sssig{.36} & \\
\bottomrule
\tablenotesx{8}{%
\textit{Note:} $N=87$\quad
\tnotex{\pnote} $p<.05$\quad
\tnotex{\ppnote} $p<.01$\quad
\tnotex{\pppnote} $p<.001$
}
\label{tab:correlations}
\end{tabularx}
\end{table}

\section{Qualitative Results}\label{sec:results-qual}

The questionnaire contained open-ended text questions to investigate the participants' general experience with passwordless authentication as well as its strengths and weaknesses.
We also included an open-ended question to understand better why participants do (not) want to adopt passwordless authentication for specific account types.
In summary, our participants' general impression was \textquote{good} (87\% P, 81\% R) or \textquote{good, but} (13\% P, 19\% R), with only two participants in \groupR not describing a positive general experience.
Based on our participants' comments, \autoref{sec:discussion-strenghts} summarizes the strengths of FIDO2 (\autoref{tab:strengths}), and \autoref{sec:discussion-weaknesses} reports its weaknesses (\autoref{tab:weaknesses}).
We also quote from our participants' free-text responses to better present the details of their reasoning.

\subsection{FIDO2 Strengths}\label{sec:discussion-strenghts}
Overall, our participants reacted positively to platform and roaming authentication on smartphones. The \gls{SUS} scores for both groups were concentrated at the top of the scale ($> 90$), indicating \emph{excellent} usability \cite{bangor2009determining}. The acceptance scores in both groups were high but significantly higher in \groupP, albeit with a small effect size.

In our lab study, participants gained hands-on experience with passwordless authentication on our mockup website.
However, we also asked them if they wanted to use the authentication methods on their own accounts, encouraging the participants to reflect on their everyday authentication use cases and whether they would like to use passwordless authentication beyond this user study.
There was a clear positive trend for each of the 11 account types, with the majority wanting to adopt passwordless authentication. The median share of positive responses (\textquote{likely} and \textquote{very likely} to adopt) across all account types was 69.8\%.
When we asked users about their general experience with the studied authentication methods, most participants (86\%) had a positive impression.

Why did users like the studied authentication methods? The coded responses to the open-ended text questions revealed four main strengths of passwordless authentication related to security, passwords, usability, and authentication times (\autoref{tab:strengths}).

\briefSection{Secure}
Most participants in both groups
commented on the methods' security benefits, e.g., one participant wrote:

\studyquote{From my point of view, it feels more secure than, for example, a password manager.}{\groupR}{32}

This is encouraging, as for any new authentication method to succeed, it must not only be secure, but users must also perceive it as secure and trust it \cite{lyastani2020kingslayer}.

\briefSection{Password Replacement}
Users generally dislike passwords and the associated overhead of creating, memorizing, and updating them \cite{ur2015added}.
Many participants stated eliminating this effort as an advantage of passwordless methods.

\briefSection{Usable}
Participants in both groups praised the usability of the studied authentication methods and described them as \textquote{intuitive}, \textquote{simple}, and \textquote{efficient}:

\studyquote{Easy to use, even for non-technical people.}{\groupP}{44}

\briefSection{Fast}
Participants, mainly in \groupP
but also in \groupR,
mentioned that the authentication method is fast.

\subsection{FIDO2 Weaknesses}\label{sec:discussion-weaknesses}

The median share of negative responses (\textquote{unlikely} or \textquote{very unlikely} adoption) across all account types was 17.6\%.
Why do some participants not want to adopt FIDO2 authentication on their smartphones, despite security benefits and excellent usability scores?
To illustrate this, note that those participants still had a respectable median \gls{SUS} score of 93.8, demonstrating that usability alone is insufficient for users to adopt FIDO2 in their everyday lives.
Instead, there are further adoption barriers, which we now present in more detail.
We start with problems common to both platform and roaming authentication and then turn to each of them individually.
\autoref{tab:weaknesses} summarizes all weaknesses mentioned by our participants within a smartphone environment.

\subsubsection{Platform and Roaming Weaknesses}\label{sec:plat-roam-weaknesses}
Overall, in both \groupP and \groupR, the usability of registration and login on the smartphone was fine, but some participants raised concerns about account recovery and delegation, authenticator revocation, and availability.
We identified the following weaknesses in both groups, indicating that these weaknesses are common to both platform and roaming authentication.

\briefSection{Account Recovery}
When users lose access to their accounts, they need a convenient and reliable recovery mechanism. Participants from both groups raised questions about how to proceed in case they suddenly cannot access their accounts:

\studyquote{If I lose the YubiKey, I will have to look for a replacement, block my accounts, and then register initially with my new YubiKey...}{\groupR}{20}

\briefSection{Authenticator Revocation}
The problem of revocation closely relates to recovery. When the smartphone or the security key is stolen, users not only need to recover access, but they also want to revoke the stolen authenticator to prevent abuse. This problem applies to both platform and roaming authentication, but participants from \groupP rarely mentioned revocation as an issue.

\begin{table}[t]
\centering
\caption{Strengths of passwordless authentication reported by our participants.}
\tablefontsize
\begin{tabularx}{\columnwidth}{@{} lXrr @{}}
\toprule
&
\tableheadline{Strength} & 
\tableheadline{\groupP} &
\tableheadline{\groupR} \\
\midrule
\rotmultirow{7}{Both} & Secure & \heatmapcell{69} & \heatmapcell{69} \\
& No password memorization issues & \heatmapcell{47} & \heatmapcell{52} \\
& Easy to use/Intuitive & \heatmapcell{38} & \heatmapcell{33} \\
& No passwords & \heatmapcell{16} & \heatmapcell{14} \\
& Easier than passwords & \heatmapcell{16} & \heatmapcell{12} \\
& No password creation & \heatmapcell{7} & \heatmapcell{5} \\
& No password updates & \heatmapcell{4} & \heatmapcell{2} \\
\midrule
\rotmultirow{5}{Platform} & Fast & \heatmapcell{47} & \heatmapcell{12} \\
& Always available & \heatmapcell{11} & \heatmapcell{2} \\
& Privacy & \heatmapcell{11} & \heatmapcell{2} \\
& Easy to setup & \heatmapcell{4} & \heatmapcell{2} \\
\\
\midrule
\rotmultirow{5}{Roaming} & One solution for many accounts & \heatmapcell{4} & \heatmapcell{10} \\
& Good for lay users & \heatmapcell{2} & \heatmapcell{5} \\
& Security less reliant on smartphone & \heatmapcell{0} & \heatmapcell{2} \\
\\ \\
\bottomrule
\tablenotesx{4}{%
\textit{Note:} Numbers in \tableheadline{\groupP} and \tableheadline{\groupR} are in-group percentages (\%). As both groups mostly identify similar strengths of passwordless authentication, \autoref{sec:discussion-strenghts} reports them jointly and points out differences.
}
\end{tabularx}
\label{tab:strengths}
\end{table}
\begin{table}[t]
\centering
\caption{Weaknesses of passwordless authentication reported by our participants.}
\tablefontsize
\begin{tabularx}{\columnwidth}{@{} lXrr @{}}
\toprule
&
\tableheadline{Weakness} &
\tableheadline{\groupP} &
\tableheadline{\groupR} \\
\midrule
\rotmultirow{7}{Both} 
& Revocation/Recovery & \heatmapcell{11} & \heatmapcell{26} \\
& Complicated for lay users & \heatmapcell{13} & \heatmapcell{5} \\
& Privacy concerns & \heatmapcell{4} & \heatmapcell{10} \\
& Website compatibility & \heatmapcell{4} & \heatmapcell{2} \\
& Coerced authentication & \heatmapcell{0} & \heatmapcell{5} \\
& Account sharing & \heatmapcell{4} & \heatmapcell{0} \\
& Unfamiliarity & \heatmapcell{4} & \heatmapcell{0} \\
\midrule
\rotmultirow{5}{Platform} & Use on multiple clients & \heatmapcell{27} & \heatmapcell{5} \\
& Technical problems & \heatmapcell{20} & \heatmapcell{5} \\
& Technology mistrust & \heatmapcell{13} & \heatmapcell{0} \\
& Empty battery & \heatmapcell{7} & \heatmapcell{0} \\
& Biometric security & \heatmapcell{4} & \heatmapcell{0} \\
\midrule
\rotmultirow{5}{Roaming} & Loss/Destruction & \heatmapcell{20} & \heatmapcell{81} \\
& Something to carry & \heatmapcell{2} & \heatmapcell{36} \\
& Theft & \heatmapcell{4} & \heatmapcell{21} \\
& Costs & \heatmapcell{0} & \heatmapcell{12} \\
& Cumbersome & \heatmapcell{2} & \heatmapcell{7} \\
\bottomrule
\tablenotesx{4}{%
\textit{Note:} Numbers in \tableheadline{\groupP} and \tableheadline{\groupR} are in-group percentages (\%).
}
\end{tabularx}
\label{tab:weaknesses}
\end{table}

\briefSection{Account Delegation}
Users are familiar with sharing access to some of their accounts, e.g., streaming services.
Furthermore, some users would like to give their partner access to a shared bank account. Account delegation is easy with passwords, but with passwordless authentication (especially platform authentication), there is nothing to share:

\studyquote{In case another person needs to log in to your account for whatever reason that's difficult, or if it's an account that multiple people use [...]}{\groupR}{34}

\briefSection{Availability}
Password-based authentication is available as long as the user remembers their password.
In contrast, passwordless authentication can be temporarily unavailable, e.g., when a user forgets to carry their YubiKey (\groupR) or the smartphone's battery is empty (\groupP).
Participants of both groups worried about the potential inconvenience of authentication being unavailable.

\subsubsection{Platform Weaknesses}\label{ssec:platform-weaknesses}

Even though our study reports good usability of platform authentication, some participants wondered how to access their accounts from other devices. Others were concerned about hardware issues or did not trust the technology.

\briefSection{Multiple Clients}
While passwords are applicable on most client devices, our participants did not know how to access their accounts on devices other than their smartphones, e.g., their \gls{PC} or laptop.
For example, one participant wrote:

\studyquote{Authentication is bound to the device. For me, it is important that these accounts are accessible from everywhere [...]}{\groupP}{57}

Similarly, some participants wondered how to access their accounts on public computers:

\studyquote{Not usable on all devices, e.g., the PC in the library.}{\groupP}{73}

\briefSection{Malfunctions} 
Participants were concerned about technical problems with the biometric sensor, preventing them from accessing their accounts. Especially participants who had used biometric authentication before to access their smartphones seemed to be aware of potential problems:

\studyquote{Sometimes the fingerprint does not work so well when the hands are wet, for example (from my own experience with unlocking the iPhone).}{\groupP}{71}

\briefSection{Technology Mistrust}
A few participants were skeptical of platform authentication due to its novelty. They were unsure if they could trust it, as Touch ID felt like a  black box to them.
For instance, one participant wrote:

\studyquote{Less visibility into what's actually happening. It feels like you're giving up a lot of control.}{\groupP}{56}

\subsubsection{Roaming Weaknesses}\label{ssec:roaming-weaknesses}

Most of our participants gave good \gls{SUS} scores to roaming authentication during our lab study, indicating that the general authentication flow is usable. However, our participants raised concerns about the requirement to purchase and carry additional hardware, which could get lost, destroyed, or stolen in the worst case.

\briefSection{Additional Hardware}
Roaming authentication requires additional hardware, which users must remember to carry. The associated physical effort is a hidden usability penalty \cite{fomichev2021fastzip}.
While we provided all hardware to our participants in our study, users would need to bring their own authenticator in practice, which requires thought and effort.

\briefSection{Loss/Destruction/Theft}
The roaming authenticator is often a small external device, so it can get lost, destroyed, or stolen. The immediate consequence is a lack of availability because users lose access to their accounts
. Most participants in \groupR (81\%) mentioned this as an issue, for example:

\studyquote{You must always have the [YubiKey] with you $\to$ can be lost, forgotten $\to$ no login possible. Actually, you always want to have it with you, but then you can lose it, in which case you can no longer log in.}{\groupR}{80}

\briefSection{Deployment Costs}
Another problem with requiring additional hardware is that it costs money. The \gls{NFC}-capable FIDO2 roaming authenticators used for our study cost 55 USD at the time of our study \cite{yubico2022yubikey5cNFC}, but cheaper models are available for 25 USD \cite{yubico2022securitykeyNFC}. Nevertheless, some users are unwilling to pay this much for authentication
, especially considering that the best practice is buying a second token as a backup.

\studyquote{The [YubiKey] costs money, to begin with, and passwords are free. I think this will prevent many users from using it.}{\groupR}{1}

For P34, however, the costs were no dealbreaker:

\studyquote{Has a certain price (but would be worth it to me personally).}{\groupR}{34}

\section{Discussion}\label{sec:discussion}
We now discuss the main findings of our study.
In \autoref{sec:discussion-accounts}, we discuss usage patterns in our participants' responses that lead to account-specific adoption decisions.
\autoref{sec:discussion-addressing} revisits the weaknesses of FIDO2 and guides on how to alleviate major issues.
Finally, we address the limitations of our work (\autoref{sec:discuss-limitations}) and discuss future work (\autoref{sec:discuss-outlook}).

\subsection{Adoption Depends on Account Type}\label{sec:discussion-accounts}

Most weaknesses presented in the previous section were only mentioned by a fraction of our participants, emphasizing their positive general impression and indicating that some of these weaknesses only apply to specific usage patterns or accounts.
After the participants stated their likelihood of using passwordless authentication for real-world online accounts, we asked them to explain what affected their decision.
\autoref{tab:adoption-reasons} and \autoref{tab:adoption-barriers} summarize the characteristics of accounts for which participants (did not) want to use passwordless authentication.
We observe that specific account characteristics affect how participants weigh the strengths and weaknesses of passwordless authentication.
Our goal is to identify account characteristics that make an account more suitable for either platform or roaming authentication.

\subsubsection{Account Sensitivity}

We observe opposite trends regarding how users reason about adopting passwordless authentication for sensitive accounts.

\briefSection{Security vs. Availability}
Some users reject roaming authentication for non-sensitive accounts, where availability concerns outweigh security benefits.
These participants cited account sensitivity as a reason for adopting passwordless authentication but its absence as a reason against adoption:

\studyquote{In principle, I would choose to use the authentication method for most accounts, except for accounts [...] to which I do not assign any importance.}{\groupR}{32} 

\briefSection{Convenience vs. Mistrust}
We identify a countertrend of users rejecting passwordless authentication for sensitive accounts, as they cannot put aside their technology mistrust in favor of the usability benefits.
These participants cited \textquote{easy to use} and \textquote{fast} as strengths of passwordless authentication and generally considered it to be \textquote{secure}.
However, they would instead rely on password-based authentication for sensitive accounts, trusting their memory more than the mechanisms of passwordless authentication:

\studyquote{I consider the [Yubikey] to be secure. However, for important accounts, I would rather rely on my memory and that the login does not get into the wrong hands.}{\groupR}{68}

\subsubsection{Usage Frequency}
During our study, there was a clear relationship between the adoption likelihood and the usage frequency of an account.

\briefSection{Security vs. Physical Effort}
Some users reject roaming authentication for frequently-used accounts because the physical effort outweighs the security benefits.
We find that these participants described the YubiKey as \textquote{easy to use} and \textquote{secure}, but also as \textquote{something to carry}:

\studyquote{For: Accounts that I do not use daily. [...]
Not: For Insta and Co. Since I use them several times a day (even on the side or so), and the effort to always use the USB stick is not in proportion to the benefit, i.e., the danger if someone steals my password or hacks my account.}{\groupR}{80}

\briefSection{Speed vs. Initial Effort}
We observe the opposite trend for platform authentication:
Some users reject platform authentication for rarely-used accounts because the effort to set up passwordless authentication outweighs the reduced authentication times.
This trend is likely related to our previous observation that more participants perceived platform authentication as fast compared to roaming authentication:

\studyquote{For: accounts where you have to log in more often (often on the same device). Against: social networks: login [only once]}{\groupP}{72}

\begin{table}[t]
    \centering
    \caption{Account characteristics favoring adoption.}
    \tablefontsize
    \begin{tabularx}{\columnwidth}{XR{30pt}R{30pt}}
    \toprule
    \tableheadline{Yes, for \dots} &
    \tableheadline{\groupP} &
    \tableheadline{\groupR} \\
    \midrule
    sensitive accounts & \heatmapcell{16} & \heatmapcell{38} \\
    rarely used accounts & \heatmapcell{0} & \heatmapcell{7} \\
    accounts that are not shared & \heatmapcell{4} & \heatmapcell{2} \\
    frequently used accounts & \heatmapcell{4} & \heatmapcell{2} \\
    business-related accounts & \heatmapcell{2} & \heatmapcell{2} \\
    accounts where fast access is crucial & \heatmapcell{4} & \heatmapcell{0} \\
    non-sensitive accounts & \heatmapcell{2} & \heatmapcell{0} \\
    accounts mainly used on smartphone & \heatmapcell{2} & \heatmapcell{0} \\
    accounts not used on other devices & \heatmapcell{2} & \heatmapcell{0} \\
    \bottomrule
    \tablenotesx{3}{%
    \textit{Note:} Numbers in \tableheadline{\groupP} and \tableheadline{\groupR} are in-group percentages (\%).
    }   
    \end{tabularx}
    \label{tab:adoption-reasons}
\end{table}

\subsubsection{Mobility}

We observe that some users reject passwordless authentication for accounts used outside of their homes because, in these scenarios, the chances of authentication unavailability increase.
These participants worried about an \textquote{empty battery} (\groupP) or described the need to remember to carry the roaming authenticator (\groupR):

\studyquote{For streaming services, I may log in from somewhere else than home, and then, if I don't have the [YubiKey] with me, I won't be able to log in.}{\groupR}{30}

\subsection{Roadmap to Addressing the Weaknesses of FIDO2}\label{sec:discussion-addressing}

We now discuss the weaknesses identified in our study and propose how to address them while considering the findings of previous user studies in different passwordless authentication scenarios (\autoref{sec:background-related-work}).
Some weaknesses, including account recovery, authenticator revocation, and account delegation, affect both platform and roaming authentication, but other issues directly stem from the concrete authentication mode.

\subsubsection{Account Recovery}
Recovery is an open problem in FIDO2 \cite{lyastani2020kingslayer,farke2020afterAll,owens2021user,rasmussen2021usability}, which the standards currently do not sufficiently address.
The best practice is to register multiple authenticators per account and keep one as a backup, consolidating the issues of acquisition costs and additional hardware.
However, this does not scale.
Furthermore, our analysis shows that users need clear instructions on dealing with recovery as they did not realize that mitigations known from password-based authentication can also work for FIDO2.
For instance, online services can allow users to request instructions for authentication reset via email and enroll new FIDO2 credentials, similar to a password reset form.
The usability and acceptance of this form of recovery require further research.
Some works propose to improve recovery in FIDO2 with extensions \cite{frymann2020asynchronous, yubico2022recovery,putz2021futureproof}, certificates \cite{conners2022letsAuthenticate}, or electronic IDs \cite{schwarz2022feido}.

\subsubsection{Authenticator Revocation}
Authenticator revocation has been identified as a weakness in prior studies of FIDO2 roaming authentication \cite{lyastani2020kingslayer,farke2020afterAll,owens2021user,rasmussen2021usability} but not in related studies of platform authentication.
A possible explanation is that there already exist mitigations to deal with the loss of smartphones.
For instance, platform authentication often requires local authentication, making it harder for attackers to access accounts with stolen smartphones.
Additionally, users can remotely erase the data on a lost iPhone \cite{apple2020remoteErase}, further addressing revocation.
Most roaming security keys do not support local authentication, although some models are equipped with biometric protection \cite{yubico2022yubikeyBio,feitian2022biopass}.
However, as the FIDO2 standards do not sufficiently solve authenticator revocation, web services (i.e., FIDO2 relying parties) need to provide account management options to revoke access from specific authenticators.

\begin{table}[t]
    \centering
    \caption{Account characteristics preventing adoption.}
    \tablefontsize
    \begin{tabularx}{\columnwidth}{XR{30pt}R{30pt}}
    \toprule
    \tableheadline{No, for \dots} &
    \tableheadline{\groupP} &
    \tableheadline{\groupR} \\
    \midrule
    non-sensitive accounts & \heatmapcell{11} & \heatmapcell{21} \\
    sensitive accounts & \heatmapcell{11} & \heatmapcell{10} \\
    shared accounts & \heatmapcell{7} & \heatmapcell{10} \\
    frequently used accounts & \heatmapcell{0} & \heatmapcell{10} \\
    accounts used outside of home & \heatmapcell{4} & \heatmapcell{5} \\
    business-related accounts & \heatmapcell{7} & \heatmapcell{2} \\
    rarely used accounts & \heatmapcell{7} & \heatmapcell{2} \\
    accounts supposed to be anonymous & \heatmapcell{0} & \heatmapcell{2} \\
    accounts used on multiple devices & \heatmapcell{2} & \heatmapcell{0} \\
    \bottomrule
    \tablenotesx{3}{%
    \textit{Note:} Numbers in \tableheadline{\groupP} and \tableheadline{\groupR} are in-group percentages (\%).
    }
    \end{tabularx}
    \label{tab:adoption-barriers}
\end{table}

\subsubsection{Account Delegation}
While giving your authenticator to someone else is possible, it only allows account delegation to one person at a time and does not work remotely \cite{lyastani2020kingslayer, rasmussen2021usability, lassak2021mitigating}.
As FIDO2 supports the registration of multiple authenticators for a single account \cite{standard2021webauthnLevel2}, we argue that account delegation can become a strength of FIDO2 rather than a weakness once online services start using this feature.

\subsubsection{Platform Authentication}
Our study confirms that the main weakness of platform authentication is the use on multiple clients \cite{lassak2021mitigating}.
While FIDO2 allows users to add multiple authenticators per account,
the individual registration of all devices for each account imposes an additional burden on users.
Furthermore, it does not work in special authentication environments, e.g., public computers in a library where roaming authentication would be better suited.

Platform authentication requires trusting the involved hardware components to handle the key material.
Even if the involved protocols and components are proven secure, they are often implemented as blackboxes, and some \glspl{TPM} had vulnerabilities in the past \cite{han2018bad,nemec2017return}.
Many users mistrust such novel, unfamiliar technologies \cite{lyastani2020kingslayer,farke2020afterAll,owens2021user,rasmussen2021usability,lassak2021mitigating,boogaard2022user}, which is a dealbreaker for the acceptance of an authentication method.
One potential solution is better education on how platform authentication, FIDO2, and public-key-based authentication methods work.

\subsubsection{Roaming Authentication}
The primary weaknesses of roaming authentication are by design:
Roaming authenticators require additional hardware that users have to purchase \cite{bonneau2012quest, lyastani2020kingslayer, farke2020afterAll},
they have to be carried around \cite{lyastani2020kingslayer,farke2020afterAll,owens2021user,rasmussen2021usability}, and they can be lost, destructed and stolen \cite{lyastani2020kingslayer, farke2020afterAll, owens2021user, rasmussen2021usability, lassak2021mitigating}.
These weaknesses can also apply to smartphones as physical devices, but more participants in \groupR described these issues, indicating that the perceived risk is higher for roaming authentication.
Furthermore, smartphones can mitigate these issues, which is infeasible with the limited capabilities of roaming authenticators:
Some smartphone vendors allow users to locate their smartphones, such as via Apple Find My \cite{apple2022findMy} or Android Find My Device \cite{google2022findMyDevice}, which mitigates the consequences of loss and theft.
Additionally, platform authentication does not cause additional deployment costs, as it runs directly on the user's smartphone, which most users already own \cite{pew2021smartphones}.

\subsection{Limitations}\label{sec:discuss-limitations}
This section addresses limitations of our work as a result of our recruitment and study design.

\subsubsection{Task Instruction}
Our instructional videos ensured that all participants had a basic understanding of passwordless authentication, resulting in a study sample that is likely better informed about FIDO2 than today's general public.
This limits the ecological validity of our results for the first-time user experience of users unaware of passwordless authentication.
A previous user study by Lassak et al. showed, however, that basic user training can address such misconceptions of first-time users \cite{lassak2021mitigating}.
In contrast, usability problems in the day-to-day user interaction of FIDO2 cannot be solved with education alone and are thus a more interesting scenario for our study.
We argue that briefing our participants supports studying the usability of FIDO2 for informed lay users, which is increasingly representative given the ongoing proliferation of FIDO2 \cite{mdn2022webauthnCompatibility}.
The instructional videos affected \groupP and \groupR in the same way, as both of our study groups watched identical videos except for the details of platform and roaming authentication.
Thus, our participants all had identical conditions, allowing for fair comparison in our between-groups design.

\subsubsection{Recruitment}
Our sample was comparatively young, with a larger-than-average share of students and most participants in the 20-39 age range.
Furthermore, not compensating participants for their participation in our study may have introduced a skew towards participants who are wealthy enough to participate in lab studies for free.
Overall, our sample shows a slight deviation from the average population in terms of demographics but, as we argue, without significant effect on the measured items.
Rather than selectively sampling participants to achieve a representative distribution, we instead controlled for factors potentially affecting usability and acceptance (\autoref{sec:control-variables}).
Additionally, our instructional videos helped balance prior knowledge, further mitigating the effects of our unfiltered sample.
Neither our control variables nor age correlated significantly with usability or acceptance.
While our sample only includes 10 participants (11\%) older than 50, there have been related usability studies focusing on older adults' (older than 60) interaction with roaming authentication \cite{das2019towards}.

\subsubsection{Mobile Operating System}
We conducted our study on Apple iOS, which at the time of our study represented the most popular mobile operating system in North America \cite{statista2022shareMobileOS}.
As the FIDO2 user interaction is the same on iOS and Android, except for the wording and illustration in the web browser prompts in Safari and Chrome, we argue that our results also generalize to Android smartphones.
Furthermore, none of the strengths, weaknesses, or adoption decisions stated by our participants directly relate to iOS but are equally relevant to FIDO2 on Android.

\subsection{Outlook and Future Work}\label{sec:discuss-outlook}
Before concluding our work, we dare a look into the future of FIDO2 and provide recommendations for future research, as well as for the FIDO Alliance and FIDO2 users.

\subsubsection{User Education}
Introducing our participants to the basics of passwordless authentication and the setup of FIDO2 might have led to improved usability compared to related studies, which should encourage the FIDO Alliance and website operators to strive for better FIDO2 education for the general public.
Nevertheless, we found that our participants described several weaknesses that are actually solvable with mechanisms of the FIDO2 standards, e.g., account delegation or authenticator reset.
The FIDO Alliance should prioritize explicitly informing users about these mechanisms of FIDO2 to mitigate such misconceptions \cite{lassak2021mitigating}, e.g., they could publish videos similar to our instructional videos (\autoref{ssec:study-design}).

\subsubsection{Compatibility}
Our study shows that users consider a variety of client devices (e.g., laptops, smartphones, smart TVs, game consoles, wearables) when reflecting on their authentication use cases.
As a consequence of our work, future authentication methods should be compatible with versatile authentication environments to qualify for the users' needs.

\subsubsection{Platform vs. Roaming}
Proper consideration of other FIDO2 weaknesses requires adding technical solutions to the FIDO2 standards or developing improved authenticators.
For future research and industry, we encourage improving platform authentication because its main weaknesses (\autoref{ssec:platform-weaknesses}) are more easily addressable than those of roaming authenticators (\autoref{ssec:roaming-weaknesses}).

Similarly, we suggest that relying parties (i.e., website operators) expand support for FIDO2 authentication. All modern smartphones support FIDO2 platform authentication nowadays, so FIDO2 is a viable second factor in addition to passwords or even a suitable single factor replacing passwords. Users can benefit from FIDO2's easy-to-use and secure authentication flow without too much overhead for website operators, as FIDO2's browser API is straightforward to implement. By relying on platform or roaming authentication, website operators also avoid storing users' passwords, eliminating a potential liability in case of data breaches.

A combination of smartphone platform authentication with the ideas of using smartphones as roaming authenticators for external devices \cite{owens2020smartphones,rasmussen2021usability} would be promising to address most weaknesses identified in this study.
For instance, Apple has announced \emph{passkeys} for iOS 16 \cite{apple2022ios16}, replacing Touch ID for the Web as the iPhone's FIDO2 platform authenticator \cite{apple2022passkeys}.

Passkeys are FIDO2 key credentials that are synchronized between the user's Apple devices.
The usability results of our study regarding FIDO2 platform authentication with Touch ID also apply to passkeys.
Furthermore, passkeys address several platform authentication-related weaknesses identified for Touch ID, e.g., allowing the use of passkeys on multiple clients and enabling users to authenticate on other Apple devices (by connecting their own Apple device as a roaming authenticator).
Additionally, the synchronization of passkeys in the iCloud Keychain simplifies FIDO2 account recovery.
While passkeys are a promising solution within the Apple environment,
the FIDO Alliance should strive for an open system that is usable on all platforms independent of a single vendor's infrastructure.

\section{Conclusion}
We conducted a between-groups lab study (N=87) of FIDO2 passwordless authentication, comparing roaming and platform authentication on smartphones.
Our main goal was to identify the strengths and weaknesses of passwordless authentication on smartphones as perceived by lay users, focusing on account-specific adoption barriers.
Our key findings for each research question are as follows:

\briefSection{RQ1}
\emph{"What is the usability and acceptance of FIDO2 passwordless authentication using platform and roaming authentication on smartphones?"}
Both platform and roaming authentication show the potential of satisfactory day-to-day usability on smartphones, but less common events such as device malfunctions, account recovery, and account delegation impair the experience. Overall, users slightly prefer platform authentication.

\briefSection{RQ2}
\emph{"What benefits and concerns do users consider when using FIDO2 passwordless authentication on smartphones?"}
Users appreciate platform and roaming authentication as simple and secure password replacements without the overhead of memorizing and managing passwords for each account.
The primary weakness identified by users is the loss/theft/destruction of the authenticator and the associated burden of revoking and recovering access to each account.
For roaming authentication, users criticize having to carry an additional device.
In contrast, for platform authentication, users are concerned with accessing their accounts on other client devices and technical problems with the biometric sensor.

\briefSection{RQ3}
\emph{"Which account types do users want to secure using FIDO2 passwordless authentication on smartphones?"}
While most users are likely to adopt passwordless authentication for their accounts, users prioritize usability, security, and availability differently depending on the account type.
As a result, the weaknesses of passwordless authentication turn into dealbreakers for specific account types and usage patterns.
Participants generally preferred platform over roaming authentication for non-sensitive accounts or frequently used accounts due to the excellent availability and the fast authentication time of the smartphone's integrated platform authenticator.
Conversely, participants preferred roaming authentication for sensitive or rarely used accounts.
Some users reject passwordless authentication for shared accounts as they do not see an option to delegate access to other users.
Overall, most users prefer passwordless authentication for sensitive accounts.
However, despite the usability benefits, some do not fully trust the technology and would therefore only use it for non-sensitive accounts.

\medskip\noindent
In summary, although there is no one-size-fits-all authenticator for all account types, we recommend improving platform authentication, which has more easily addressable weaknesses than roaming authentication.

\begin{acks}
We thank the anonymous reviewers for their helpful suggestions.
Furthermore, we thank Annemarie Mattmann for her support during the qualitative data analysis, and Jasin Machkour for the helpful discussions on robust data analysis. 
This work has been co-funded by the LOEWE initiative (Hesse, Germany) within the emergenCITY center and the Federal Ministry of Education and Research of Germany in the project Open6GHub (grant number: 16KISK014).
\end{acks}

\section*{Availability}

Together with this paper, we release a replication package with our evaluation scripts and the pseudonymized dataset generated in our study \cite{seemoo2023zenodo} consisting of usability and acceptance scores, the adoption likelihood for 11 account types, and 9 control variables for each of our 87 participants.
We also provide our mockup website's source code \cite{seemoo2023github} to facilitate future work.

\bibliographystyle{ACM-Reference-Format}
\bibliography{bibliography}

\appendix

\section{Demographics}\label{sec:demographics}

In this section, we extend our participants' descriptive data (\autoref{tab:descriptive}) with further demographic information (\autoref{tab:demographics}).

\begin{table}[!hb]
\centering
\caption{%
Our participants' demographic data.
\typofix{We report in-group percentages (and frequencies), Pearson's chi-square test, and Cram\'er's V as the effect size (ES)}.
}
\tablefontsize
\setlength{\tabcolsep}{4pt}
\begin{tabularx}{\columnwidth}{Xlrrcr}
\toprule
&
\tableheadline{Variable} &
\tableheadline{\groupP} &
\tableheadline{\groupR} &
\tableheadline{Statistic} &
\tableheadline{ES}\\
&& ($N=45$) & ($N=42$) \\
\midrule
\multirow{4}*{Gender} & Female & 44.4 (20) & 28.6 (12) & \multirow{4}*{\shortstack{$\chi^2(1)=2.35$ \\ $p=.125$}} & \multirow{4}*{.16} \\
& Male & 55.6 (25) & 71.4 (30) && \\
& Other & 0 (0) & 0 (0) & & \\
& No answer & 0 (0) & 0 (0) & & \\
\midrule
\multirow{6}*{Age} & 18-19 & 6.7 (3) & 7.1 (3) & \multirow{6}*{\shortstack{$\chi^2(5)=4.94$ \\ $p=.423$}} & \multirow{6}*{.11} \\
& 20-29 & 48.9 (22) & 69.0 (29) &&\\
& 30-39 & 28.9 (13) & 14.3 (6) &&\\
& 40-49 & 2.2 (1) & 0 (0) &&\\
& 50-59 & 11.1 (5) & 7.1 (3) &&\\
& 60-69 & 2.2 (1) & 2.4 (1) &&\\
\midrule
\multirow{7}*{Education} & Still in school & 4.4 (2) & 0 (0) & \multirow{7}*{\shortstack{$\chi^2(6)=3.89$ \\ $p=.692$}} & \multirow{7}*{.09} \\
& Middle school & 8.9 (4) & 9.5 (4) &&\\
& High school & 37.8 (17) & 26.2 (11) &&\\
& Bachelor & 35.6 (16) & 45.2 (19) &&\\
& Master & 8.9 (4) & 11.9 (5)  &&\\
& Doctorate & 2.2 (1) & 2.4 (1)  &&\\
& Other & 2.2 (1) & 4.8 (2) &&\\
\bottomrule
\end{tabularx}
\label{tab:demographics}
\end{table}

\section{Questionnaire}\label{sec:questionnaire}
We provide a translation of our questionnaire, with descriptive question names for convenience.
The actual questionnaire only had non-descriptive section names ("Category 1" and so on) so as to not influence the participants.
We gave the questionnaire to the participants in the native language of the country where we ran the study.

\briefSection{Usability}
\begin{task}
Please mark the answer that reflects your immediate response to each statement.
Please do not think too long about each statement and be sure to provide an answer to all statements.
\end{task}

\rating{Five responses ranging from \textquote{Strongly disagree} to \textquote{Strongly agree}}

\begin{category}
    \qitem{1} I think that I would like to use this system frequently.
    \qitem{2} I found the system unnecessarily complex.
    \qitem{3} I thought the system was easy to use.
    \qitem{4} I think that I would need the support of a technical person to be able to use this system.
    \qitem{5} I found the various functions in this system were well integrated.
    \qitem{6} I thought there was too much inconsistency in this system.
    \qitem{7} I would imagine that most people would learn to use this system very quickly.
    \qitem{8} I found the system very cumbersome to use.
    \qitem{9} I felt very confident using the system.
    \qitem{10} I needed to learn a lot of things before i could get going with this system.
\end{category}

\briefSection{Acceptance}
\begin{task}
Now please evaluate the system.
To do this, read each pair of words carefully and make one cross per line.
\end{task}

\rating{Five \typofix{responses} in between the words}

\begin{category}
    \qitem{1} Useless vs. Useful
    \qitem{2} Pleasant vs. \typofix{Unpleasant}
    \qitem{3} Bad vs. Good
    \qitem{4} Nice vs. Annoying
    \qitem{5} Effective vs. Superfluous
    \qitem{6} Irritating vs. Likeable
    \qitem{7} Assisting vs. Worthless
    \qitem{8} Undesirable vs. Desirable
    \qitem{9} Raising Alertness vs. Sleep-Inducing
\end{category}

\briefSection{Privacy Concerns}
\begin{task}
Please indicate how strongly you agree with the following statements
\end{task}

\rating{Seven responses ranging from \textquote{Strongly disagree} to \textquote{Strongly agree}}

\begin{category}
    \qitem{1} I am concerned that companies are collecting too much personal information about me.
    \qitem{2} I am concerned about my privacy.
    \qitem{3} To me it is important to keep my privacy intact.
    \qitem{4} Novel technologies are threatening privacy increasingly.
\end{category}

\briefSection{Technology Affinity}
\begin{task}
In the following questionnaire, we will ask you about your interaction with technical systems.
The term “technical systems” refers to apps and other software applications, as well as entire digital devices (e.g., mobile phone, computer, TV, car navigation).
\end{task}

\rating{Six responses from \textquote{Completely disagree} to \textquote{Completely agree}}

\begin{category}
    \qitem{1} I like to occupy myself in greater detail with technical systems.
    \qitem{2} I like testing the functions of new technical systems.
    \qitem{3} I predominantly deal with technical systems because I have to.
    \qitem{4} When I have a new technical system in front of me I try it out intensively.
    \qitem{5} I enjoy spending time becoming acquainted with a new technical system.
    \qitem{6} It is enough for me that a technical system works; I don't care how or why.
    \qitem{7} I try to understand how a technical system exactly works.
    \qitem{8} It is enough for me to know the basic functions of a technical system.
    \qitem{9} I try to make full use of the capabilities of a technical system.
\end{category}

\briefSection{iOS Familiarity}
\begin{category}
    \qitem{1} Do you use an iPhone for private or business purposes? \rating{Yes / No}
\end{category}

\briefSection{Open-Ended Questions}
\begin{category}
    \qitem{1} How would you describe your general experience with the presented authentication method? \rating{Free text response}
    \qitem{2} Which advantages do you see in the usage of the presented \typofix{authentication} method? \rating{Free text response}
    \qitem{3} Which disadvantages do you see in the usage of the presented authentication method? \rating{Free text response}
\end{category}

\briefSection{Adoption}
\begin{task}
In the following, think of a website or account that you use yourself. Assume that the service to which this account belongs supports the presented authentication method. How likely is it that you would use the presented authentication method for this account?
\end{task}

\rating{For items 1-11, there are five responses ranging from \textquote{very unlikely} to \textquote{very likely}, with an additional response for \textquote{Not available}}

\begin{category}
    \qitem{1} Social network account
    \qitem{2} Streaming service account
    \qitem{3} Messenger service account
    \qitem{4} Travel portal account
    \qitem{5} Gaming portal account
    \qitem{6} Internet service provider account
    \qitem{7} Google/Apple account
    \qitem{8} Private email account
    \qitem{9} Business email account
    \qitem{10} Online shop account
    \qitem{11} Online banking account
    \qitem{12}  Please briefly explain for which accounts you would decide to use or not use the presented authentication method. \rating{Free text response}
\end{category}

\briefSection{2FA Familiarity}
\begin{task}
Which two-factor authentication methods have you previously used?
\end{task}

\begin{category}
    \qitem{1} Text messages \rating{Checkbox}
    \qitem{2} TAN lists \rating{Checkbox}
    \qitem{3} Code generators \rating{Checkbox}
    \qitem{4} Smartphone apps \rating{Checkbox}
    \qitem{5} Hardware keys \rating{Checkbox}
    \qitem{6} Other: \rating{Free text response}
\end{category}

\briefSection{Demographics}
\begin{category}
    \qitem{1} Please state your gender. \rating{Male / Female / No answer / Other}
    \qitem{2} How old are you? \rating{10-19 / 20-29 / 30-39 / 40-49 / 50-59 / 60-69 / 70-79 / $\geq$80}
    \qitem{3} Please state your highest educational degree: \rating{Still in school / Middle school / High school / Bachelor's degree / Master's degree or Diploma / Doctorate / Other}
    \qitem{4} Please state your area of work or area of studies. \rating{Free text response}
    \qitem{5} Is there anything else you would like to tell us? \rating{Free text response}
\end{category}

\section{Codebooks}\label{sec:codebooks}

This section lists the codebooks (\autoref{tab:codebooks}) that we created during the qualitative analysis of the four open-ended text-questions.

\aptLtoX[graphic=no,type=env]{\begin{table}
\centering
\caption{Codebooks created during the qualitative analysis (\autoref{sec:methods-analysis}) of the open-ended text questions.}
\label{tab:codebooks}
\tablefontsize
\begin{tabular}{L{0.23\columnwidth}L{0.77\columnwidth}}
    \hline
\multicolumn{2}{c}{(a) Strengths} \\\midrule
    Usability &
        \codex{Easy to use/Intuitive}
    	\codex{Good for lay users}
    	\codex{Fast}
    	
    	\codex{Easy to setup}
    \\\hline
    Availability &
    	\codex{One solution for many accounts}
        \codex{Always available}
    \\\hline
    Cognitive effort &
    	\codex{No passwords}
        \codex{No password creation}
        
    	\codex{No password updates}
    	\codex{Easier than passwords}
    	
    	\codex{No password memorization issues}
    \\\hline
    Security &
        \codex{Secure}
    	\codex{Security less reliant on smartphone}
    	\codex{Privacy}
    	\codex{Security less reliant on website}
    \\\hline
\multicolumn{2}{c}{(b) Weaknesses} \\\midrule
    User experience &
        \codex{Slow}
    	\codex{Complicated for lay users}
    	\codex{Cumbersome}
    \\\hline
    Deployability &
        \codex{Costs}
    	\codex{Website compatibility}
	\\\hline
    Availability &
        \codex{Technical problems}
    	\codex{Something to carry}
    	
    	\codex{Empty battery}
    	\codex{Loss/ Destruction}
    \\\hline
    Mental models &
        \codex{Unfamiliarity}
    	\codex{Revocation/Recovery}
    	
    	\codex{Account sharing}
    	\codex{Technology mistrust}
    	
    	\codex{Use on multiple clients}
    \\\hline
    Security &
        \codex{Theft}
    	\codex{Coerced authentication}
    	\codex{Biometric security}
    	\codex{Privacy concerns}
    \\\hline
\multicolumn{2}{c}{(c) General Expression} \\\midrule
    Conclusion &
        \codex{Good}
        \codex{Good, but}
        \codex{Bad, but}
        \codex{Bad}
    \\\hline
    Usability &
        \codex{Easy to use / Understandable}
        \codex{Cumbersome}
        \codex{Fast}
        
        \codex{Complicated Safari UI}
    \\\hline
    Security &
        \codex{Secure}
        \codex{Too easy, can feel insecure}
        
        \codex{Privacy concerns}
    \\\hline
    Cognitive effort &
        \codex{No password memorization issues}
        
        \codex{Easier than a password manager}
        
        \codex{(Needs) no passwords}
    \\\hline
    Availability &
        \codex{Forgetting/Loss/Theft/Destruction}
    \\\hline
    Mental models &
        \codex{Technology mistrust}
        \codex{Important}
        \codex{Innovative}
        
        \codex{More transparency/ information required}
    \\\hline
\multicolumn{2}{c}{(d) Adoption reasons} \\\midrule
    Importance &
        \codex{Sensitive accounts}
    	\codex{Non-sensitive accounts}
    \\\hline
    Business Stuff &
        \codex{Business-related accounts}
    \\\hline
    Anonymity &
        \codex{Accounts supposed to be anonymous}
    \\\hline
    Multiple Devices &
    	\codex{Accounts not used on other devices}
    	
        \codex{Accounts mainly used on smartphone}
        
    	\codex{Accounts used on other devices}
    \\\hline
    Account Sharing &
        \codex{Shared accounts}
    	\codex{Accounts that are not shared}
    \\\hline
    Frequency &
        \codex{Rarely used accounts}
    	\codex{Frequently used accounts}
    \\\hline
    Speed &
    	\codex{Accounts where fast access is not important}
        
        \codex{Accounts where fast access is crucial}
    \\\hline
    Mobility &
        \codex{Accounts used outside of home}
    \\\hline
\end{tabular}
\end{table}}{\begin{table}
\centering
\caption{Codebooks created during the qualitative analysis (\autoref{sec:methods-analysis}) of the open-ended text questions.}
\label{tab:codebooks}
\tablefontsize
\begin{tabular}{L{0.23\columnwidth}L{0.77\columnwidth}}
    \toprule
\rowcolor{\altrowcolor}\multicolumn{2}{c}{\tableheadline{(a) Strengths}} \\\midrule
    Usability &
        \codex{Easy to use/Intuitive}
    	\codex{Good for lay users}
    	\codex{Fast}
    	
    	\codex{Easy to setup}
    \\\midrule
    Availability &
    	\codex{One solution for many accounts}
        \codex{Always available}
    \\\midrule
    Cognitive effort &
    	\codex{No passwords}
        \codex{No password creation}
        
    	\codex{No password updates}
    	\codex{Easier than passwords}
    	
    	\codex{No password memorization issues}
    \\\midrule
    Security &
        \codex{Secure}
    	\codex{Security less reliant on smartphone}
    	\codex{Privacy}
    	\codex{Security less reliant on website}
    \\\midrule
\rowcolor{\altrowcolor}\multicolumn{2}{c}{\tableheadline{(b) Weaknesses}} \\\midrule
    User experience &
        \codex{Slow}
    	\codex{Complicated for lay users}
    	\codex{Cumbersome}
    \\\midrule
    Deployability &
        \codex{Costs}
    	\codex{Website compatibility}
	\\\midrule
    Availability &
        \codex{Technical problems}
    	\codex{Something to carry}
    	
    	\codex{Empty battery}
    	\codex{Loss/ Destruction}
    \\\midrule
    Mental models &
        \codex{Unfamiliarity}
    	\codex{Revocation/Recovery}
    	
    	\codex{Account sharing}
    	\codex{Technology mistrust}
    	
    	\codex{Use on multiple clients}
    \\\midrule
    Security &
        \codex{Theft}
    	\codex{Coerced authentication}
    	\codex{Biometric security}
    	\codex{Privacy concerns}
    \\\midrule
\rowcolor{\altrowcolor}\multicolumn{2}{c}{\tableheadline{(c) General Expression}} \\\midrule
    Conclusion &
        \codex{Good}
        \codex{Good, but}
        \codex{Bad, but}
        \codex{Bad}
    \\\midrule
    Usability &
        \codex{Easy to use / Understandable}
        \codex{Cumbersome}
        \codex{Fast}
        
        \codex{Complicated Safari UI}
    \\\midrule
    Security &
        \codex{Secure}
        \codex{Too easy, can feel insecure}
        
        \codex{Privacy concerns}
    \\\midrule
    Cognitive effort &
        \codex{No password memorization issues}
        
        \codex{Easier than a password manager}
        
        \codex{(Needs) no passwords}
    \\\midrule
    Availability &
        \codex{Forgetting/Loss/Theft/Destruction}
    \\\midrule
    Mental models &
        \codex{Technology mistrust}
        \codex{Important}
        \codex{Innovative}
        
        \codex{More transparency/ information required}
    \\\midrule
\rowcolor{\altrowcolor}\multicolumn{2}{c}{\tableheadline{(d) Adoption reasons}} \\\midrule
    Importance &
        \codex{Sensitive accounts}
    	\codex{Non-sensitive accounts}
    \\\midrule
    Business Stuff &
        \codex{Business-related accounts}
    \\\midrule
    Anonymity &
        \codex{Accounts supposed to be anonymous}
    \\\midrule
    Multiple Devices &
    	\codex{Accounts not used on other devices}
    	
        \codex{Accounts mainly used on smartphone}
        
    	\codex{Accounts used on other devices}
    \\\midrule
    Account Sharing &
        \codex{Shared accounts}
    	\codex{Accounts that are not shared}
    \\\midrule
    Frequency &
        \codex{Rarely used accounts}
    	\codex{Frequently used accounts}
    \\\midrule
    Speed &
    	\codex{Accounts where fast access is not important}
        
        \codex{Accounts where fast access is crucial}
    \\\midrule
    Mobility &
        \codex{Accounts used outside of home}
    \\\bottomrule
\end{tabular}
\end{table}}

\end{document}